\documentstyle [12pt]{article}
\newcommand{\pa}{\partial}
\newcommand{\al}{\alpha}
\newcommand{\be}{\beta}
\newcommand{\de}{\delta}
\newcommand{\De}{\Delta}

\newcommand{\ga}{\gamma}
\newcommand{\Ga}{\Gamma}
\newcommand{\la}{\lambda}
\newcommand{\La}{\Lambda}
\newcommand{\si}{\sigma}
\newcommand{\vep}{\varepsilon}
\newcommand{\vp}{\varphi}

\newcommand{\eq}{\begin{equation}}
\newcommand{\eqe}{\end{equation}}
\newcommand{\qed}{\hfill \rule {1ex}{1ex}\\ }

\begin{document}
\message{reelletc.tex (Version 1.0): Befehle zur Darstellung |R  |N, Aufruf=
z.B. \string\bbbr}
%
%
%  Sonderzeichen
\message{reelletc.tex (Version 1.0): Befehle zur Darstellung |R  |N, Aufruf=
z.B. \string\bbbr}
\font \smallescriptscriptfont = cmr5
\font \smallescriptfont       = cmr5 at 7pt
\font \smalletextfont         = cmr5 at 10pt
\font \tensans                = cmss10
\font \fivesans               = cmss10 at 5pt
\font \sixsans                = cmss10 at 6pt
\font \sevensans              = cmss10 at 7pt
\font \ninesans               = cmss10 at 9pt
\newfam\sansfam
\textfont\sansfam=\tensans\scriptfont\sansfam=\sevensans
\scriptscriptfont\sansfam=\fivesans
\def\sans{\fam\sansfam\tensans}
%----------------------------------------------------------
\def\bbbr{{\rm I\!R}} %reelle Zahlen
\def\bbbn{{\rm I\!N}} %natuerliche Zahlen
\def\bbbE{{\rm I\!E}} %Einheitsmatrix by I. Zoller
\def\bbbm{{\rm I\!M}}
\def\bbbh{{\rm I\!H}}
\def\bbbk{{\rm I\!K}}
\def\bbbd{{\rm I\!D}}
\def\bbbp{{\rm I\!P}}
\def\bbbone{{\mathchoice {\rm 1\mskip-4mu l} {\rm 1\mskip-4mu l}
{\rm 1\mskip-4.5mu l} {\rm 1\mskip-5mu l}}}
\def\bbbc{{\mathchoice {\setbox0=\hbox{$\displaystyle\rm C$}\hbox{\hbox
to0pt{\kern0.4\wd0\vrule height0.9\ht0\hss}\box0}}
{\setbox0=\hbox{$\textstyle\rm C$}\hbox{\hbox
to0pt{\kern0.4\wd0\vrule height0.9\ht0\hss}\box0}}
{\setbox0=\hbox{$\scriptstyle\rm C$}\hbox{\hbox
to0pt{\kern0.4\wd0\vrule height0.9\ht0\hss}\box0}}
{\setbox0=\hbox{$\scriptscriptstyle\rm C$}\hbox{\hbox
to0pt{\kern0.4\wd0\vrule height0.9\ht0\hss}\box0}}}}

\def\bbbe{{\mathchoice {\setbox0=\hbox{\smalletextfont e}\hbox{\raise
0.1\ht0\hbox to0pt{\kern0.4\wd0\vrule width0.3pt height0.7\ht0\hss}\box0}}
{\setbox0=\hbox{\smalletextfont e}\hbox{\raise
0.1\ht0\hbox to0pt{\kern0.4\wd0\vrule width0.3pt height0.7\ht0\hss}\box0}}
{\setbox0=\hbox{\smallescriptfont e}\hbox{\raise
0.1\ht0\hbox to0pt{\kern0.5\wd0\vrule width0.2pt height0.7\ht0\hss}\box0}}
{\setbox0=\hbox{\smallescriptscriptfont e}\hbox{\raise
0.1\ht0\hbox to0pt{\kern0.4\wd0\vrule width0.2pt height0.7\ht0\hss}\box0}}}}

\def\bbbq{{\mathchoice {\setbox0=\hbox{$\displaystyle\rm Q$}\hbox{\raise
0.15\ht0\hbox to0pt{\kern0.4\wd0\vrule height0.8\ht0\hss}\box0}}
{\setbox0=\hbox{$\textstyle\rm Q$}\hbox{\raise
0.15\ht0\hbox to0pt{\kern0.4\wd0\vrule height0.8\ht0\hss}\box0}}
{\setbox0=\hbox{$\scriptstyle\rm Q$}\hbox{\raise
0.15\ht0\hbox to0pt{\kern0.4\wd0\vrule height0.7\ht0\hss}\box0}}
{\setbox0=\hbox{$\scriptscriptstyle\rm Q$}\hbox{\raise
0.15\ht0\hbox to0pt{\kern0.4\wd0\vrule height0.7\ht0\hss}\box0}}}}

\def\bbbt{{\mathchoice {\setbox0=\hbox{$\displaystyle\rm
T$}\hbox{\hbox to0pt{\kern0.3\wd0\vrule height0.9\ht0\hss}\box0}}
{\setbox0=\hbox{$\textstyle\rm T$}\hbox{\hbox
to0pt{\kern0.3\wd0\vrule height0.9\ht0\hss}\box0}}
{\setbox0=\hbox{$\scriptstyle\rm T$}\hbox{\hbox
to0pt{\kern0.3\wd0\vrule height0.9\ht0\hss}\box0}}
{\setbox0=\hbox{$\scriptscriptstyle\rm T$}\hbox{\hbox
to0pt{\kern0.3\wd0\vrule height0.9\ht0\hss}\box0}}}}

\def\bbbs{{\mathchoice
{\setbox0=\hbox{$\displaystyle     \rm S$}\hbox{\raise0.5\ht0\hbox
to0pt{\kern0.35\wd0\vrule height0.45\ht0\hss}\hbox
to0pt{\kern0.55\wd0\vrule height0.5\ht0\hss}\box0}}
{\setbox0=\hbox{$\textstyle        \rm S$}\hbox{\raise0.5\ht0\hbox
to0pt{\kern0.35\wd0\vrule height0.45\ht0\hss}\hbox
to0pt{\kern0.55\wd0\vrule height0.5\ht0\hss}\box0}}
{\setbox0=\hbox{$\scriptstyle      \rm S$}\hbox{\raise0.5\ht0\hbox
to0pt{\kern0.35\wd0\vrule height0.45\ht0\hss}\raise0.05\ht0\hbox
to0pt{\kern0.5\wd0\vrule height0.45\ht0\hss}\box0}}
{\setbox0=\hbox{$\scriptscriptstyle\rm S$}\hbox{\raise0.5\ht0\hbox
to0pt{\kern0.4\wd0\vrule height0.45\ht0\hss}\raise0.05\ht0\hbox
to0pt{\kern0.55\wd0\vrule height0.45\ht0\hss}\box0}}}}

\def\bbbz{{\mathchoice {\hbox{$\sans\textstyle Z\kern-0.4em Z$}}
{\hbox{$\sans\textstyle Z\kern-0.4em Z$}}
{\hbox{$\sans\scriptstyle Z\kern-0.3em Z$}}
{\hbox{$\sans\scriptscriptstyle Z\kern-0.2em Z$}}}}

\title{ Mass Generation in the large $N$-nonlinear $\sigma$-Model}

\author{Christoph Kopper\\
Centre de Physique Th{\'e}orique de l'Ecole Polytechnique\\
F-91128 Palaiseau France }  
\date{ March 15, 1998 }

\maketitle

\begin{abstract}
Abstract: We study the infrared behaviour of the two-dimensional
Euclidean $O(N)$ nonlinear $\si$-Model with a suitable ultraviolet cutoff.
It is proven that for a sufficiently large (but finite!) number $N$
of field components the model is massive and thus has 
exponentially decaying correlation functions. We use a representation
of the model with an interpolating bosonic field. This permits to
analyse the infrared behaviour
without any intermediate breaking of $O(N)$-symmetry.
The proof is simpler than that of the corresponding
result for the Gross-Neveu-Model [1].

\end{abstract}

\noindent
\section{Introduction}

We want to study the infrared behaviour of the two-dimensional
Euclidean nonlinear $\si$-model [2] which is formally given in terms of
the Lagrangian
\eq
{\cal L} \,=\, 
\frac{N}{2\la} \{\,(\pa \phi)^2\,+\,\frac{K}{4}(\phi^2-1)^2\}\,\,.  
\eqe
Here the  constant $K$ is assumed to be of order
1, whereas we assume $N>\!>1\,$, for $\la$ see below. 
$\,\phi$ is a real-valued 
$N$-(flavour-)component bosonic
field in the fundamental (vector) representation of $O(N)$. 
The minimum of $\cal L$ is thus situated at
$\phi^2\,=\,1$, where the value $1$ may be changed by rescaling the
field variable.
The ultraviolet (UV) cutoff as well as more
precise statements on the lower bound for $N$ will be specified later. 
As regards $\la$, its value should not be much larger than 1,
because otherwise the generated mass $m$ 
approaches the UV cutoff, see below (20). 
If it is much smaller than 1, on the other hand, the effective energy
range of the UV cutoff model becomes large and therefore the bounds, 
which involve factors of $\exp(4\pi/\la)$, deteriorate.
The convergence proof then requires larger values of $N$. 
In the full renormalization
group construction one would try to impose a condition $\la\,\sim
\,1$ by fixing the renormalized coupling $\la_0$ of the last
renormalization group step to obey that condition since 
in the full construction
$\la_0$ corresponds to our coupling $\la$.

The  standard nonlinear $\si$-model has the constraint on the field 
variable (which we call $\phi$ instead of $\si$ )
\eq
\phi^2\,=\,1\,\,.
\eqe
Condition (2) can be obtained from (1) by a suitable limit taking
$K\,\to\,\infty$.\footnote{For the analysis of the model in that limit 
a lattice regularization is probably most appropriate, see also [25, 26]
and the comments below.} 
Such a constraint however is immediately 
softened out when starting from the model with a large UV cutoff
on integrating out high frequency modes, even after 
the first renormalization group step in a renormalization group
construction. This can be seen from the renormalization group 
construction of the hierarchical model which has been performed by
Gawedzki and Kupiainen[4] and later also by Pordt and Reiss[5].
It is rather obvious anyhow: Once you have (at least) two independent
frequency modes, fluctuations of one may compensate those of the
other such that the constraint (2) is restored for the sum.
These fluctuations are not even highly improbable since neighbouring 
frequency
modes may look similar in position space for frequencies close to
the border line between the two.
Thus we obtain for $K$ a value of order 1 after the first step. 
The much more
difficult part of the ultraviolet analysis of the model - so far
only performed in the hierarchical case for $N>2$ and as long as the
effective coupling stays small -
is to show that the Lagrangian (1) is a good approximation 
to  the full model. That implies in particular
that the model has only one marginal
direction which is well represented by the quartic term in (1).
So our starting point is reasonable 
when giving credit to the evidence based on the hierarchical 
approximation. This hierarchical analysis in turn agrees with the seminal
papers on the model based on perturbation theory
by Br{\'e}zin, Le Guillou and Zinn-Justin [6], and
the analysis of Br{\'e}zin and Zinn-Justin [6]
also  agrees with ours on the IR side in the  limit $N \to \infty$.
Furthermore the generally accepted view 
is confirmed by
numerical simulations [7] and, which is of great importance in this
respect, also by the Bethe Ansatz methods 
based on the exact $S$-matrix [3], which
show in particular that the model has a mass gap. 
Nevertheless these results are not fully based 
on well proven assumptions and are rather
self-consistent than rigorous. So we note that on the other hand that
doubts against the general wisdom 
have been raised  by   Patrascioiu and Seiler [8].

We  take an UV regularized version of (1) as our starting point. The scale
is chosen such that the UV cutoff $\La$ is situated at $\La\,=\,1$.
The situation in constructive field theory is often complicated by the
fact that the expansions around the situation where the degrees of
freedom are to some extent decoupled 
start from regularized versions which
tend to violate symmetries of the
model in question. The symmetries on the other hand often greatly simplify the 
perturbative analysis if,  as is often the case, 
an invariant regularization for perturbation theory is at hand. 
Fortunately this time we are on the easy side:
Once we have introduced an interpolating field, which we now call
$\si$, the whole analysis of the model can be performed without 
breaking the $O(N)$-symmetry, in complete agreement with the
Mermin-Wagner-theorem [9],[10]. When the one-component scalar $\si$-field
has been introduced we may integrate out the $\phi$-field thus
obtaining a new interaction given by (the inverse of) a Fredholm determinant.
For the UV cutoff model it is well-defined in finite volume.
The infinite volume limit is taken in the end, once the cluster
and Mayer expansions have been performed, which allow to divide out the
divergent vacuum functional. The analysis of the Fredholm determinant
proceeds similarly as that of the corresponding determinant in the
case of the Gross-Neveu-Model [1]. It is 
simplified in the same way as
the expansions are since we do not have to distinguish different zones
characterized by the mean value of the $\si$-field - apart from the
small field/large field splitting. The main new problem
lies in the fact that for the inverted Fredholm determinant 
some of the estimates used to bound the determinant
(together with antisymmetric 
tensor products generated by taking derivatives when cluster expanding,
see [1],p.169 and more generally [11]) are no more valid. 
The problem is solved by 
deriving new bounds on inverted Fredholm determinants 
- in the last part of Ch.3, to show stability, 
by introducing a finer splitting of the large 
field configurations 
before cluster expanding to make sure that the cluster expansion 
derivatives always produce small terms, and by evaluating the
expansion derivatives through Cauchy formulae.

The paper is organized as follows: Our specific choices for the regulators
and the basic definitions are presented in Ch.2. They are dictated by
technical simplicity. In Ch.3 we perform the small/large field 
splitting and develop the bounds on the various terms in the action
ensuing from that splitting, as well as on the non-local operator
kernels appearing. In particular we show that all the kernels appearing
fall off exponentially in the small field region. 
In Ch.4 the cluster-expansion is performed
which then allows to control the thermodynamic limit
and to prove the exponential fall-off of the (two-point) correlation
function(s).   

After submitting this paper we learned about two important references
on the subject. First the author was not aware of Kupiainen's work
\footnote{This important and beautiful contribution to constructive 
physics is maybe
not as well known as it should be to those working in the field. In
part this might be due to its title.} 
[25]. Secondly, few weeks after submission there appeared a preprint by Ito
and Tamura [26]. We close the introduction by shortly commenting on
these papers. 
Kupiainen regards the $N$ component nonlinear $\si$-model on a unit
width lattice for arbitrary dimensions d. He shows that the
$1/N$-expansion is asymptotic above the spherical model critical
temperature $T_S$, which is zero for $d=2$. He also proves 
the existence of a mass gap for these temperatures and $N$
sufficiently large. Without attaching much importance to the numerical
side we just say what '$N$
sufficiently large ' means. We read from [25] (see equ.(19))  
for  the two-dimensional case
that for given inverse temperature $\be$ one needs
\[
N\,>cst\,e ^{50 \pi\be}. 
\]
Since $\be$ is to be identified with the inverse coupling $1/\la$ 
in our language
this is basically the same as our bound: We require
\[
N^{-1/6}\,<\,cst\, e ^{-4 \pi/\la} 
\]
since the small factor per cluster expansion step (see end of
sect.4.4) has to beat the factor $O(m^{-2})$ from the spatial
integration per link.
Similarly the authors of [26] state their result in Theorem 24 for
\[
N\,>cst\,e ^{400 \pi\be} 
\]
and $\be$ large.
They regard the same model as Kupiainen, 
the $N$ component nonlinear $\si$-model on a unit
width lattice, for $d=2$. 
Thus [25] and [26] analyse the lattice version of (1) where the limit $K\to
\infty$ has been taken, i.e. the Heisenberg model. 
The result [26] only concerns the free energy
or partition function
which is shown to be analytic in $\be$, given  $N$ as above.
Correlation functions have not yet been treated. It seems clear
however that their method of proof which, as ours, is based on a small/large
field cluster expansion is well adapted for that case too.
We prove exponential fall-off of the two-point function,
extension to any connected $n$-point function is straightforward
using the Mayer expansion formulae for those, see e.g. [19].
The change in sect.4.5 would consist in singling out a connecting tree
now for  $n$ external points instead of two. 
Kupiainen's result on the other hand 
is based on reflection positivity in the form of chess board estimates. 
It is not clear how the result on the exponential
fall-off can be extended to general connected functions  
in this context, so strictly speaking (as he does) his result only holds for
those correlation functions which have no nontrivial truncations.
\footnote{In special cases he  succeeds in
performing truncations by a clever use of certain Ward identities.} 
An important point shared by [25] and [26] (in fact the authors
of [26] could have referred themselves to [25] here) is that they
both apply the Brydges-Federbush random walk representation to show
and use
exponential fall-off of the lattice kernels of $1/[p^2+m^2+i\si]$. 
In the continuum we
only succeed in proving exponential fall-off for small fields
$\si$. This is the main reason why we introduce a whole hierarchy of
large field regions with larger and larger protection corridors
(see (60)-(63)), and the fall-off over the corridors has to make up
for the (possibly) absent fall-off in the large field domain. 
Apart from this [26] is technically closer to my 
paper than to [25].  It is more detailed on some aspects of the 
expansions. A number of bounds take similar form
here and in [26]. In [26] the building blocks of the cluster expansion
are taken to be large also in the small field region. 
This has technical advantages, on the other
hand treating many degrees of freedom as a whole 
generally tends to deteriorate the numerical bounds.

\section{Presentation and Rigorous Definition of the Regularized Model}
We want to show that the UV regularized large $N\,$ $\si$-model is
massive, i.e. that the correlation functions decay exponentially.
In our explicit representation we will restrict to the two-point
function, generalizations to arbitrary $2N$-point functions being
obvious. Thus formally we study the following object:
\eq
S_2(x,y) \,\sim\, 
\int D\vec{\phi}\,\,\phi_i(x)\phi_i(y)
e^{-\frac{N}{2\la} \int \{\,(\pa \phi)^2\,+\,\frac{K}{4}(\phi^2-1)^2\}}\,\,.  
\eqe
Here $D\vec{\phi}$ indicates the product of (ill-defined) Lebesgue measures
$D\phi_1,\ldots, D\phi_N$. 
Before giving sense to this expression mathematically 
by imposing suitable regulators we want to
introduce the interpolating field $\si$ as announced. We rewrite (3)
as 
\eq
S_2(x,y) \,\sim\, 
\int D\vec{\phi} D\si\,\,\phi_i(x)\phi_i(y)
e^{-\frac{N}{2\la} \int \{\,(\pa \phi)^2\,+\,i(\phi^2-1)\si\,+\,
\frac{1}{K}\si^2\}}\,\,  
\eqe
up to a global field-independent normalization factor. Now we can
perform the Gaussian integrations over the $\phi$-fields to obtain
\eq
S_2(x,y) \,\sim\, 
\int D\si\,(\frac{1}{p^2+i\si})(x,y)\,\,
{\det} ^{-N/2}(p^2+i\si)\,\,
e^{-\frac{N}{2\la K} \int \, \si^2\,+\,\frac{iN}{2\la}\int \si}\,\,  
\eqe
again up to a global field-independent normalization factor and on
rescaling $\phi^2 \to \phi'^2=(N/2\la)\phi^2\,$.
$\,(\frac{1}{p^2+i\si})(x,y)$ denotes the position space kernel
of the operator $\frac{1}{p^2+i\si}$. Its existence in
 ${\cal L}^2(\bbbr^2)$  say, will be clear once the cutoffs and thus the
support of the measure are specified below.\footnote{ When studying
  higher order coorrelation functions it is preferable to work in the 
space $\bigoplus_1^N {\cal L}^2(\bbbr^2)$ and to suppress the exponent
$N$ of $\det$ instead, because in this case the factor replacing 
$\,(\frac{1}{p^2+i\si})(x,y)$ will depend on the flavour indices.
We will adopt this convention only in the last part of the paper where
it somewhat shortens the notation.}
As regards {\bf notation} we will generally use the {\sl same letters for
position and momentum space objects}. This lack in precision in our
eyes is overcompensated by the gain in suggestive shortness.  
For the same reason and on the basis of the previous remarks on the
size of the constants appearing we will abbreviate by {\sl O(1)
sums of products of $N$-independent constants} the largest of which 
appearing will actually be $1/m^2$ (21). Without making this explicit 
we pay some attention not to collect astronomic numbers into O(1).

By performing a translation of the field variable $\si$ according   
to 
\eq
\tau' \,=\,\si\,+\,i m^2
\eqe
we finally arrive at
\eq
S_2(x,y) \sim 
\int \! D\tau (\frac{1}{p^2+m^2+ig\tau})(x,y)\,
{\det}^{-{N \over 2}}(1\,+\,\frac{ig}{p^2+m^2}
\tau)\,
e^{-\frac{1}{2} \int \! \tau^2\,+\,i\sqrt{N}(\frac{m^2}{\sqrt{\la K}}
+\sqrt{\frac{K}{4\la}})\int\! \tau}\,.
\eqe
This time the change of normalization  stems from three sources:
from the translation, from a change of normalization of the Fredholm
determinant and from a rescaling of the $\tau'$-field: 
$\tau'\,\to\,\tau=\sqrt{\frac{N}{\la K}}\,\tau'$. 
In (7) we introduced the coupling constant 
\eq
g\,=\, \sqrt{\frac{\la K}{N}}\,\,.
\eqe
The value of the translation parameter $m$ is fixed below ((17)-(21)) by
a gap equation. This eliminates
the term  in the interaction exponential which is linear in $\tau$,
and this in turn 
is a prerequisite in the $1/N$-expansion, since that term has 
a coefficient $\sim \sqrt{N}$.
Before specifying the UV and IR regularizations we note that
from the point of view of mathematical purity it would have been
preferable to introduce them from the beginning. This however
would have blown up the previous manipulations without a real gain
since (3) and (7) are in fact to be viewed on equal footing as starting points:
They both produce the same perturbation theory in $1/N$.\\
We now introduce the following regularizations:\\
{\bf UV1}: We set the cutoff scale to be $1$ and replace
\eq
p^2 \,\to \, p^2_{reg}\,=\,p^2\,e^{p^2} 
\eqe
in (7).\\
{\bf UV2}: We also introduce an UV cutoff for the $\tau$-field.
When tracing this back to the original interaction (1) it amounts 
to smoothing out the pointlike quartic $\phi^4$-interaction. 
To the expression $\,\int D\tau \, e^{-\frac{1}{2} \int \, \tau^2}$
in (7) corresponds in rigorous notation 
integration with respect to the Gaussian measure $d\mu_{\de}(\tau)$ 
with mean zero and covariance $C_{\de}(x-y)\,=\,\de(x-y)$, or in
momentum space $C_{\de}(p)\,=\,1(p)$. We replace the $\de$-function by a
regularized version
\eq
1(p)\,\to\,\frac{1}{1+\hat{f}(p)}\,,\quad\hat{f}(p)\,=\,\sqrt{1+\pi(p)}\,f(p)\,
\sqrt{1+\pi(p)}\,,
\eqe
where $\pi(p)$ is defined below, see (28). It is a smooth 
nonnegative function depending
on $p^2$ only, bounded above by a constant of order $1/m^2$ (see (29)).
$\,f(p)$ also is a smooth nonnegative function depending on $p^2$
only. It vanishes in the origin, grows monotonically with $p^2$ such that
$\,\alpha (p^2)^2 < f(p) < A (p^2)^2$ with suitable $0<\,\alpha< A
\,<\,\infty$,
and fulfills 
\eq
(\frac{1}{1+f})(x-y)\,=\,0 \,,\,\,\mbox{ if}\, |x-y|>1. 
\eqe
The last condition is the most important one.
That all conditions are mutually compatible is rather credible.
A proof is
in the elementary Lemma 1 in [1].\footnote{ In fact we did not prove 
monotonicity in [1]. This can however be achieved by a slight
extension of the proof. We do not include it since monotonicity is not
needed here, it might however be useful when performing a
renormalization group construction on the same basis.}
There a suitable $f$ (which in [1] is
further restricted by demanding that it should vanish of high order 
in the origin) is constructed explicitly,
basically by starting from the characteristic function of the unit ball 
in position space $\bbbr^2$ and taking linear combinations of rescaled 
convolutions thereof. 
We should note that it is by no means crucial to have a cutoff with
these particular properties. Only sufficient fall-off of $\,\,1/(1+f)\,$
in momentum and position space are required. So e.g. $1/(1+e^{p^2})$
would do. The compact support property (11) is however helpful when
fixing the final covariance of the model, taking into account large
field constraints, see (67). It
eliminates further small correction terms of similar nature as those   
appearing in $\de C_{\ga}$ (70), cf. the remark after (79).
In short the UV cutoff on the $\tau$ field replaces the ultralocal
covariance $\de(x-y)$ of this field by a smoothed compact support
version of the $\de$-function sandwiched between the two 
$\sqrt{1+\pi}$-factors. The growth properties of $f(p)$
restrict the support properties of the corresponding Gaussian measure
$d\mu_f(\tau)$ to (real) continuous functions [12] and therefore we need not
regularize expressions such as $\tau^2$ etc.
In general we will view $\tau$ as an element of the real Hilbert space
${\cal L}^2(\bbbr^2,\bbbr)$.\\
{\bf IR}: As an intermediate IR regularization to be taken away in the
end we also introduce a finite volume cutoff. To be definite we
choose a square 
\eq
\La \subset \bbbr^2
\eqe
centered at the origin with volume 
\eq
|\La|\,=\,4n^2>\!>1,\quad n\in \bbbn\,.
\eqe
We then restrict the support of the $\tau$-field to $\La$. But we do not
restrict the Gaussian measure to $\La$ from the beginning, because this
again would increase the number of correction terms later when we 
perform a configuration dependent  change of covariance. We want to
avoid this, but nevertheless want to suppress contributions in the
measure supported outside $\La$. We therefore introduce a term
\eq
\exp(-R \int_{\bbbr^2-\La}\,\tau ^2)\,\,,\qquad R>>1
\eqe
in the functional integral, and take the limit $R\to \infty$ later on.
Note that absorbing this term in the measure and taking the limit
right away would amount to restricting the covariance to $\La$ from the
beginning [13].  Again our particular
choices for the IR cutoff are convenient, but not crucial.

With these preparations we now obtain the following 
rigorous expression for the regularized normalized two-point function:\\
\eq 
S^{\La}_2(x,y) \,=\, \frac{1}{\hat{Z}^{\La}}\,
\int d\mu_f(\tau)\,(\frac{1}{p_{reg}^2+m^2+ig\tau \chi_{\La}})(x,y)\,
\eqe
\[
\times\,\,{\det}^{-N/2}(1\,+\,\frac{1}{p_{reg}^2+m^2}
ig\tau\chi_{\La})\,e ^{-R \int_{\bbbr^2-\La}\,\tau ^2}\,\,
e^{\,i\sqrt{N}(\frac{m^2}{\sqrt{\la K}}
+\sqrt{\frac{K}{4\la}})\int_{\La} \tau}\,\,.
\]
The partition function $\hat{Z}^{\La}$ is given by 
\eq 
\hat{Z}^{\La}\,=\, 
\int d\mu_f(\tau)\,{\det}^{-N/2}(1\,+\,\frac{1}{p_{reg}^2+m^2}
ig\tau\chi_{\La})\,e ^{-R \int_{\bbbr^2-\La}\,\tau ^2}\,\,
e^{\,i\sqrt{N}(\frac{m^2}{\sqrt{\la K}}
+\sqrt{\frac{K}{4\la}})\int_{\La} \tau}\,\,.
\eqe
Here $\chi_{\La}$ is the sharp characteristic function of the set
$\La$ in position space. 
Instead of $\chi_X$ we will mostly use $P_X$ to denote the 
orthogonal projector on the subspace of functions supported in $X$.
From the bounds on  the action is given in the next section
it is clear that $\hat{Z}^{\La}$
will not vanish in finite volume (see (118)). 
In the following we will mostly {\it suppress explicit reference to
the regulators}
by $reg$ and $\chi_{\La}$ for shortness.\\

As announced the value of $m$ is fixed by imposing a gap
equation eliminating the linear term in $\tau$ from the action,
i.e. we demand:
\eq
i\sqrt{N}(\frac{m^2}{\sqrt{\la K}}
+\sqrt{\frac{K}{4\la}})\int_{\La} \tau\,=\,
N/2 \,Tr(\frac{1}{p_{reg}^2+m^2}ig\,\tau \chi_{\La})\,\,.
\eqe
When evaluating the $Tr$, the term $\int_{\La}\tau$ factorizes an both
sides of (17), and we obtain the relation 
\eq
1/2\,\int \frac{d^2p}{(2\pi)^2}\,\frac{1}{p^2e^{p^2}\,+\,m^2}\,=\,
\frac{m^2}{\la  K}\,+\, \frac{1}{2\la}\,.
\eqe
For a sharp cutoff at $p^2=1$ we would find from this
\eq
\frac{1}{4\pi}\ln(\frac{1+\,m^2}{m^2})\,=\,1/\la\,+\,2\frac{m^2}{\la K} 
\eqe
with the solution
\eq
m^2\,=\,
e^{-\frac{4\pi}{\la}}\,(1+O(\frac{4\pi}{\la K}e^{-\frac{4\pi}{\la}}))\,.
\eqe
For the case of an exponential cutoff the integral cannot
be evaluated analytically, but it is easy to find suitable
upper and lower bounds saying that
\eq
m^2\,=\,c_m \,e^{- \frac{4\pi}{\la}}\,,
\eqe
where the constant $c_m$ is close to one (lies between 0.9 and 1.1)
for $\la \leq 1$. For definiteness we will assume from now on
\eq
2/\pi \,<\,\la \,<\, \pi \,\,\,\mbox{ so that }\,\,\, 
e ^{-10} \,<\,m\,<\,1/6\,\,.
\eqe
Taking into account the constraint (17) we thus obtain for the two-point
function
\eq 
S^{\La}_2(x,y) = \frac{1}{\hat{Z}^{\La}}
\int d\mu_f(\tau)\,(\frac{1}{p^2+m^2+ig\tau })(x,y)
{\det}_2^{-N/2}(1\,+\,\frac{1}{p^2+m^2}
ig\tau\chi)\, e ^{-R \int_{\bbbr^2-\La}\tau ^2}
\eqe
where we used the standard definition 
\eq
{\det}_{n+1}(1+K)\,=\,\det(1+K)\,e^{-TrK+{1\over 2}TrK^2+\ldots 
+(-1)^{n}{1 \over n} TrK^n} 
\eqe
for any traceclass operator K and $n\in \bbbn$.
In an expansion based on the parameter $1/N$ the canonical choice of
covariance is such that it contains all terms of the action quadratic  
in the field $\tau$, possibly up to terms which are suppressed for 
$N \to \infty$. This is not
yet the case for (23) since the term quadratic in $\tau$ from $\det\,$ is not
suppressed for $N$ large: It contains $1/N$ from $g^2$ and N from
${\det}^{-N/2}$ giving $N^0$ altogether.
Thus the appropriate presentation of the two-point function is rather
\eq 
S^{\La}_2(x,y)=\frac{1}{Z^{\La}}
\int d\mu_C(\tau)(\frac{1}{p^2+m^2+ig\tau})(x,y)
{\det}_3^{-N/2}(1\,+\,\frac{1}{p_{reg}^2+m^2}
ig\tau)\,e ^{-R \int_{\bbbr^2-\La}\tau ^2}\,.
\eqe
A corresponding change of definition has also been introduced when
passing from $\hat{Z}^{\La}$ to $Z^{\La}$
\eq
Z^{\La}\,=\,
\int d\mu_C(\tau)\,
{\det}_3^{-N/2}(1\,+\,\frac{1}{p_{reg}^2+m^2}
ig\,\tau\,\chi_{\La})\,e ^{-R \int_{\bbbr^2-\La}\tau ^2}\,.
\eqe
In (25), (26)
$d\mu_C(\tau)$ represents the Gaussian measure with covariance
\eq
C\,=\,(1\,+\,\hat{f}\,+\,P_{\La}\pi P_{\La} )^{-1}\,.
\eqe
$P_{\La}$ is the orthogonal projector onto the subspace
${\cal L}^2(\La)$ of ${\cal L}^2(\bbbr^2)$,
and  $\pi$ is the quadratic part in $\tau$ from $\det$.
In momentum space it is given  as
\eq
\pi(p)\,=\,
\frac{\la K}{2}\,\int\frac{d^2q}{(2\pi)^2}\frac{1}{q^2\,+\,m^2}
\frac{1}{(p+q)^2\,+\,m^2}\,>0\,.
\eqe
Since the integral is UV convergent, it is largely independent
of the cutoff functions $e^{p^2}$ which we did not write explicitly.
We find in particular
\eq
\pi(0)\,=\, \frac{C_{\pi}}{8 \pi}
\frac{\la K}{m^2} \,, 
\eqe
where $C_{\pi}$ is again a constant close to $1$. Furthermore
one easily realizes that 
\eq
\pi \leq \pi(0) \,\,\mbox{ in the operator sense,$\,$ or
in momentum space }\,\,\,\,
\pi(0)-\pi(p)\geq 0\,\,.
\eqe 
This can either be done by direct calculation or by noting that
\eq
(\tau,\pi(0)\, \tau)\,=\frac{N}{2}\,Tr(V\,V^*)\,\geq \,\frac{N}{2}
TrV^2\,=\,(\tau,\pi
\,\tau)\,,
\eqe
for the operator
\eq
V(\tau)\,=\, \frac{1}{p^2\,+\,m^2}\,g\,\tau\,.
\eqe
By $(\tau,\pi\,\tau)$ we denote the scalar product, which is given by  
$\int \tau(x) \,\pi(x-y) \,\tau(y)$ in position space.
Note that $V$ has real expectation values in the real Hilbert space
${\cal L}^2(\La,\bbbr)$.\\
For later use we collect the following facts about the operator
$\,\pi$ and (some functions of) the kernels of
$\,\pi$ and of $\,1/(p^2\,e^{p^2}+m^2)$ in position space. \\[.1cm]
{\bf Lemma 1:}\\
a) The operator $\,\pi$ fulfills: $ 0\,\leq \,\pi \,\leq \pi(0)\,\,$.\\ 
b)The kernel of $\pi$ in position space denoted
as $\pi(x-y)$ (using translation invariance) satisfies:\\
i) $|\pi(x-y)|\,\leq\,O(1)\, e^{-2m|x-y|}\,$,\\
ii) furthermore $|\sqrt{1+\pi}^{\pm 1}(x-y)|\,\leq\,O(1)\,e^{-2m|x-y|}\,$
for $x \,\neq \, y$.\\
c) $0\,<\,[1/(p^2\,e^{p^2}+m^2)](x-y)\,< \,O(1)\exp\{-m|x-y|\}$\,.\\[.1cm]
{\sl Proof:} The proof of a) was given previously. The statement b)i) 
follows from standard analyticity arguments: $\pi(p)$ is analytic in
momentum space for $(Im p)^2 \leq 4m^2$ as is directly seen from the
integrand in (28) by shifting the integration variable $q$ by $p/2$. 
The main reason to choose the analytic regulator function $e^{p^2}$
was that it does preserve (and even slightly enlarge)\footnote{by a
factor of $|O(1)m^2|$, see also [1]} 
this analyticity domain so that b)i)
follows. Coming now to the statement in b)ii)
we first note that the condition $x\neq y$ eliminates the $\de$-distribution
contribution to the kernel so that we may regard in fact
\eq
(\sqrt{1+\pi})^{\pm 1}\,-\, 1
\eqe
As compared to b)i) we now also have to verify that 
the real part of $1+\pi$ stays positive for $(Im p)^2 \leq 4m^2$
(so as to exclude a cut, i.e. a violation of analyticity due to the
square root). Again explicit calculation  simply reveals this
to be the case, where the regularization again slightly improves the 
situation. \\
c) This statement was proven in [1], Lemma 5. The lower bound follows from
the representations 
\eq
\frac{1}{p^2\,e^{p^2}+m^2}\,=\,
e^{-p^2}\frac{1}{p^2+1} \sum_{n=0}^{\infty}
(\frac{1}{p^2+1}\,-\,m^2\frac{1}{p^2+1}e^{-p^2})^n
\eqe
and
\eq
\frac{1}{p^2+1}\,-\,m^2\frac{1}{p^2+1}e^{-p^2}\,=\,
\frac{1}{p^2+1}(1-m^2 e)\,+\,m^2 e\,\int_0^1 ds \,e^{-s(p^2+1)}\,.
\eqe
Since $m^2 e\, <\,1$ it becomes now obvious by explicit calculation 
of the Fourier transforms that the kernel of
$\frac{1}{p^2\,e^{p^2}+m^2}$ is pointwise positive.
\footnote{This fact will be useful later (see in particular (173)),
but it is not crucial.}\qed

\section{Small/Large Field Decomposition and Bounds} 
The representation of the correlation functions according to (25), (26) 
is well-suited for an expansion in $1/N$, since the remnants of the
action  left in ${\det}_3$ are all suppressed by factors of $1/\sqrt{N}$
or smaller. We then have to bound
the contributions from 
${\det}_3$ for large values of the field variable $\tau$
to show that it is integrable with respect to the Gaussian measure
$d\mu_C(\tau)$. From our starting point
we presume that this should be possible, 
since there the action was manifestly integrable.
However, to obtain a convergent expansion of the
correlation functions we have to perform a cluster expansion which
makes visible the decoupling of the degrees of freedom 
with increasing separation in space. The cluster expansion
interpolation formulae 
modify all nonlocal kernels of the theory, the modification   
being  different for the measure and ${\det}_3$. Therefore one
 global bound is not sufficient. 
What we rather need are local bounds per degree of freedom.
The solution we adopt is similar as in [1], with simplifications  due to
the fact that we only have one phase, and complications  
due to the fact 
that the model is not fermionic in
origin. The latter  implies that certain sign cancellations due to the Pauli
principle are absent in the outcome of the cluster expansion and
necessitates finer distinctions on the size of the $\tau$-field
than in [1]. 
 
We distinguish between small and (a series of) large field
configurations depending on the size of $\int_{\De}\tau^2$, where 
$\De$ is any (closed) unit square in $\La$ 
with integer valued lower left corner coordinates $(n_1,n_2)$. 
Then we sum over the possible choices for all squares. 
For a given configuration we  take the union 
of large field squares, enlarge this region by adding all squares   
below some finite distance from those and divide 
(roughly speaking) the enlarged region
into its connected components. In the interior of any such component
we do not introduce interpolation parameters, it is even reasonable  
not to absorb the quadratic part of ${\det}_2$
in the covariance there. Rather we use the large field criteria
and certain bounds on inverted Fredholm determinants to show that
these regions are suppressed in probability per large field square
$\De$ and according to the size
of $\int_{\De}\tau^2$. Then the expansion is largely restricted to the
small field region,
where the integrability of ${\det}_3$
is assured due to the small field criterion anyway.
As usual such a cluster expansion with constraints goes in hand with
a certain amount of combinatorics and technicalities coming from all
sorts of correction terms. These are controlled by means of the large
value of $N$. We are now going to make this reasoning precise.

We subdivide the volume $\La$ into the  $4n^2$ unit squares $\De$
specified above
and regard some given $\tau \in {\cal L}^2(\La)$. We say that 
$\De \in \La$ is a large field square w.r.t. $\tau$ if
\eq
\la K\,\int_{\De} \tau^2 \,\geq \,N^{1/6}\,,
\eqe
and  $\De \in \La$ is a small field square w.r.t. $\tau$ if
\eq
\la K\,\int_{\De} \tau^2 \,< \,N^{1/6}\,.
\eqe
We introduce a smoothed monotonic step function 
$\theta \in C^{\infty}(\bbbr)$  
fulfilling
\eq
\theta(x)\,= \,
\left \{ \begin{array}{r@{\quad,\quad}l} 
0 & \quad \mbox{for}
\quad x \leq -1/4    \\  1
 & \quad\mbox{for} \quad x\geq 1/4\\
\end{array}  \right.
\eqe
Then we also introduce $4n^2$ factors of $1$ into the functional integral
according to 
\eq
1_{\De}\,=\,\theta(\frac{\la K ||\tau_{\De}||^2_2}{N^{1/6}}\,-\,1)\,+\,
(1\,-\,\theta(\frac{\la K ||\tau_{\De}||^2_2}{N^{1/6}}\,-\,1))\,\,.
\eqe
In (39)  we set as usual
\eq
||\tau_{\De}||^2_2\,=\,\int_{\De}\tau ^2\,\,.
\eqe
Now the first factor is decomposed further writing 
\eq
\theta(\frac{\la K ||\tau_{\De}||^2_2}{N^{1/6}}\,-\,1)\,=\,
\sum_{n=1}^{\infty}
\bigl[ \theta(\frac{\la K ||\tau_{\De}||^2_2}{N^{n/6}}\,-\,1)\,-\,
\theta(\frac{\la K ||\tau_{\De}||^2_2}{N^{(n+1)/6}}\,-\,1)
\bigr] =:
\sum_{n=1}^{\infty}\theta_n( ||\tau_{\De}||^2_2)\,\,.
\eqe
We then may rewrite (25), (26) as a sum of $2^{4n^2}$ terms each carrying
for any square $\De$ a factor which is either the first
or the second summand in (39). For a square 
carrying the first factor the functional integral is then split up
further according to (41). To fix the language we say\\
{\bf Definition:}
A square carrying the factor 
\eq
\theta^s_{\De}(\tau)
\,:=\,1\,-\,\theta(\frac{\la K ||\tau_{\De}||^2_2}{N^{1/6}}\,-\,1)
\eqe
is called a small field or s-square. A square $\De$
carrying the factor 
\eq
\theta^l_{\De}(\tau)
\,:=\,\theta(\frac{\la K ||\tau_{\De}||^2_2}{N^{1/6}}\,-\,1)
\eqe
is a called a large field or l-square. More specifically 
we call it an $l^n$-square if it carries a factor 
$\theta_n( ||\tau_{\De}||^2_2)$ resulting from the splitting (41).\\[.1cm]   
An $l$-square then only contributes to the functional integral if
\eq
\la K ||\tau_{\De}||^2_2\,\geq\, \frac{3}{4}\,N^{1/6}\,\,,
\eqe
an $l^n$-square only if
\eq
\frac{5}{4}\,N^{(n+1)/6}\,>\,\la K ||\tau_{\De}||^2_2
\,\geq\, \frac{3}{4}\,N^{n/6}\,\,,
\eqe
and an s-square only
contributes, if 
\eq
\la K ||\tau_{\De}||^2_2 < \frac{5}{4}\,N^{1/6}\,\,,
\eqe
so that 
{\it we will always assume the respective inequality to hold
once a square has been specified to be
$l$, $l^n$  or $s$}, since in this paper we are
only bounding contributions to the functional integral.
As regards notation we will write  
$P_s \,,\,\,P_l\,,\,\, P_{l^n}$ and $P_{\De}$ for the orthogonal projectors 
onto functions with support in $\La_s\,,\,\,\La_l\,,\,\,\La_{l^n}$ and $\De$
respectively. Here we  denote by $\La_s \subset \La$ 
resp. $\La_l \subset \La$ resp. $\La_{l^n} \subset \La$ the set of
small field resp. large field resp. $l^n-$ squares in
$\La$. Note $\La_s \cup \La_l \,=\, \La\,,
\,\,\bigcup_{n\in\bbbn}\La_{l^n}=\La_l\,$. 
Before proceeding further with the l/s decomposition we want to show
that the small field condition is sufficient to obtain a small upper
bound in norm on the operator appearing in $\det$:\\[.1cm] 
{\bf Proposition 2:} For $\tau \in {\cal L}^2(\La)$ let $\La_s \subset \La$ 
be a collection of unit squares such that for $\De \in  \La_s$ we have
\eq
\la K ||\tau_{\De}||^2_2\,<\, 5/4\,N^{1/6}\,\,.
\eqe
Then the operator norm of $A_s:{\cal L}^2(\La)\,\to\, {\cal L}^2(\La)$
satisfies:
\eq
||A_s|| \,\leq\, O(1) N^{-5/12}\,\leq\, N^{-2/5}\,\,.
\eqe
Here $A$ is the operator $P_{\La}\,
\frac{1}{p_{reg}^2+m^2}\,g\tau \,P_{\La}$, and 
$A_s$ is defined to be  $P_s\,A \,P_s$.
\\[.1cm]
{\sl Proof:}
We first regard $A_{\De}\,=\,P_{\De}AP_{\De}$ for $\De \in \La_s$.
For $\vp \in {\cal L}^2(\De)$ and $||\vp||_2\,=\,1$ we find:
\eq
|(A_{\De}\vp,A_{\De}\vp)|\,\leq\,
g^2\,\int_{x,y,z}|\vp(x)\tau(x)F(x-y)\chi_{\De}(y)F(y-z)\tau(z)\vp(z)|
\eqe
\[
\,\leq\, g^2\, F^2(0) \int_{x,y}|\vp(x)\tau(x)\tau(y)\vp(y)|
\,\leq\,g^2 \,F^2(0) \int_{x\in\De} \tau^2(x)\,<\,F^2(0)\,
{5 \over  4}\,N^{-5/6}\,\,.
\]
Here $F(x-y)$ is the pointwise positive kernel (see Lemma 1)
\eq
F(x-y)\,=\,
\int\frac{d^2q}{(2\pi)^2}\frac{e^{iq(x-y)}}{q^2e^{q^2}\,+\,m^2}\,\,\,,
\eqe
which is obviously bounded by its value at $0$, which in turn is
bounded by $O(1/\la)$, which we absorb in $O(1)$, which we bound
by $N^{-2/5+5/6}$.
This proves the assertion for a single square $\De$. To go 
from here to the general case one has to exploit the exponential
fall-off of the kernel $F(x-y)$ (Lemma 1), which deteriorates the
bound by a factor of $O(1/m^2)$, which we absorb in $O(1)$
and bound it again by $N^{-2/5+5/6}$.
So now let $\vp \in {\cal L}^2(\La)$ with $||\vp||_2\,=\,1$.
\eq
|(A_s\vp,A_s\vp)|\,\leq\,
\sum_{\De,\De',\De''\in \La_s}  |(A P_{\De}\vp,P_{\De'}A P_{\De''}\vp)|
\eqe
\[
\,\leq\,
O(1)\,g^2\sum_{\De,\De',\De''\in \La_s} 
\exp^{\{-m(dist(\De,\De ')+dist(\De ',\De ''))\}}
\,\,\int_{x,y}|\tau(x)\chi_{\De}(x)\vp(x)\tau(y) 
\chi_{\De ''}(y)\vp(y)|\,\,.
\]
By performing first the sum over $\De ''$ and then over $\De '$ and using
the bound on $\tau$, the Schwarz inequality and the fact 
that $\vp$ is normalized, we obtain the bound
\eq
O(1) N^{-1}\,N^{1/12}\sum_{\De \in \La_s} (\int_{\De}\vp^2\,)\,
(\int_{\De}\tau^2)^{1/2}
\,\leq\, O(1)N^{-5/6}\,\,.
\eqe
This ends the proof.\qed 

As announced we want - for given l/s-regions - enlarge the l-regions
by security belts of sufficient  width such that the fall-off
of the kernels from Lemma 1 will produce a small factor
if the kernels have to bridge these belts. This procedure generally will
merge together some of the different connected components of the  
l-region. Let $\La_{l}^1,\ldots , \La_{l}^r$ be the connected components of
$\La_l$. We say there is a {\it connectivity link} between 
$\La_{l}^i$ and $\La_{l}^j$, $1\leq i,j \leq r, i\neq j$, if 
there exists some $\De_i \in \La_{l}^i$ and some $\De_j \in \La_{l}^j$
such that there exists $\De \in \La$ with  
\eq
dist(\De_i,\De)\,+\,dist(\De_j,\De) \leq 2M \,,
\eqe
where we choose for definiteness
\eq 
M\,=\,\frac{2}{m}\,\ln N\,\,.
\eqe 
Then we call $l_1,\ldots,l_s$ the maximal subsets of $\La_l$
connected by connectivity links and call them {\it connectivity components}.
Obviously $s\leq r$. Now we set
\eq
\Ga\,=\,\Ga(l)\,=\,\{\De \subset \La |\,\, dist(\De,\La_l)\leq M\}
\eqe
and 
\eq
\Ga_i\,=\,
\Ga(l_i)\,=\,\{\De \subset \La |\,\, 
\La_{l}^k\subset l_i\,,\,\,\,dist(\De,\La_{l}^k)\leq M\}\,\,.
\eqe
Thus there is a one-to-one relation between 
the $\Ga_i$ and the $l_i$, and the $\Ga_i$ are connected 
\footnote{It requires some (elementary) work to really give an explicit
proof of that fact, which amounts basically to transferring the
square $\De$ constituting the connectivity link between 
$\De_i$ and $\De_j$ to the centre of a line 
of minimal length connecting 
$\De_i$ and $\De_j$ and showing that then either this 
transferred square or two of its neighbours touching each other
connect together $\De_i$ and $\De_j$ within some $\Ga_k$. 
We skip the proof since it is not crucial for us 
that the $\Ga_i$ are connected.}
(in the standard sense), and we have
\eq
\Ga_i \cap \Ga_j \,=\,0 \,\,\mbox{ for } \,
i\neq j\,,\,\,\mbox{ and } \,\,                 
\bigcup_1^n \Ga_i\,=\,\Ga \,\,.
\eqe
In set-theoretic relations we always denote by $0$ a set of
(standard) Lebesgue measure $0$.  
We also introduce the sets $\ga_i$ which (roughly speaking) lie between 
$l_i$ and $\Ga_i$~:
\eq
\ga_i\,=\,\{\De \subset\bbbr^2 |\,\,\La_{l}^k\subset l_i\,,
\,\,\,dist(\De,\La_{l}^k)\leq M/2\,\}
\,,\,\,\,\, \ga\,=\,\bigcup_1^n \ga_i
\eqe
so that 
\eq
dist(\ga,\La-\Ga)\geq \,M/2\,-\,\sqrt{2}\,\,.
\eqe
Note that for technical reasons we have defined $\ga$ as a subset of
$\bbbr^2$, not necessarily of $\La$. We do so because 
this definition of $\ga$ is useful when fixing the
covariance in the presence of large field configurations such that it has
good positivity and fall-off properties (see (67)-(70) and Lemma 3).

The previous definitions now are extended to the
situation where we split up further the $\La_l$-region into the
components $\La_{l^n}$. If the size of the field is very large 
we also need very large security belts to protect our large field
regions - such that the decay of the kernel across this belt again assures a
small contribution.  We start again from the connected components 
$\La_{l}^1,\ldots , \La_{l}^r$ of $\La_l$ and say that there is an 
{\it $e$-connectivity link} (or extended connectivity link) between 
$\La_{l}^i$ and $\La_{l}^j$, $1\leq i,j \leq r, i\neq j$, if 
there exists some $\De_i \in \La_{l}^i \cap \La_{l^{n'}}$ and 
some $\De_j \in \La_{l}^j\cap \La_{l^{n''}}$
such that there exists $\De \in \La$ with 
\eq
dist(\De_i,\De)\,+\,dist(\De_j,\De) \leq (n'\,+\,n'')M \,\,.
\eqe
The {\it $e$-connectivity components} are then the maximal subsets of
$\La_l$ connected by $e$-connecti\-vity links.     
We call them $l^e_i$, $1\leq i \leq s'$,
 and obviously $s'\leq\, s\,\leq \,r\,$.
Now we set

\eq
\Ga^e\,=\,\Ga^e(l)\,=\,
\bigcup_n\{\De \subset \La |\,\, dist(\De,\La_{l^n})\leq n\,M\}
\eqe
and 
\eq
\Ga^e_i\,=\,
\Ga ^e(l^e_i)\,=\,
\bigcup_n\{\,\De \subset \La |\,\,
\La_{l^n}^k:=\La_{l^n}\cap \La_{l}^k \subset l^e_i\,,\,\,
dist(\De,\La_{l^n}^k)\leq n\,M \,\}\,\,.
\eqe
Again there is a one-to-one relation between 
the $\Ga ^e_i$ and the $l^e_i$, and as before
\eq
\Ga ^e_i \cap \Ga ^e_j \,=\,0 \,\,\mbox{ for } \,
i\neq j\,,\,\,\mbox{ and } \,\,                 
\bigcup_1^{s'} \Ga ^e_i\,=\,\Ga ^e \,\,.
\eqe

Starting from the l/s- decomposition of the volume $\La$ we now decompose
the Fredholm determinant, define the s-dependent final covariance
and bound the large field action. With the definition of the operator
$A$ (Proposition 2) we can write the Fredholm determinant as $\det(1+iA)$.
We first separate $A_s$ from the rest of $A$ via the standard
relation
\eq
{\det}^{-1}(1+iA)\,=\, {\det}^{-1}(1+iA_s)\,{\det}^{-1}(1+\frac{1}{1+iA_s}iA'')
\eqe
with
\eq
A'':=\,A\,-\,A_s\,=\,A'\,+\,A_l\,,\quad
A':=\,P_s\,A\,P_l\,+\,P_l\,A\,P_s\,\,.
\eqe
Since $A$ has real spectrum, the operator $1/(1+iA)$ is well-defined.
For $A_s$ we now proceed as indicated before (see (25)), i.e. we
absorb the quadratic part in $\tau$ into the covariance. When doing so we
obtain  the following (transitory) expression for the inverse propagator
$C^{-1}_{ls}$:
\eq
C^{-1}_{ls}\,=\,P_s\,\pi \,P_s\,+\,1\,+\,\hat{f}\,\,.
\eqe
We express  (66)  in terms of  
$C^{-1}_{\ga}$ (67), the basic reason for this being the fact that we are
not able to deduce suitable fall-off properties in position space 
for the inverse of (66). Our final choice for the configuration
dependent covariance will rather be 
\eq
C^{-1}_{\ga}\,=\,\sqrt{1+\pi}(1\,-\,P_{\ga}\,+\,\vep\,P_{\ga}\,+\,f)
\sqrt{1+\pi}\,\,.
\eqe  
Here $\vep$ is introduced so that $C_{\ga}$ is bounded also in the
large field region. We fix it as 
\eq
\vep\,=\, N^{-\frac{2}{5}}\,\,. 
\eqe
Choosing (67) we have to control the difference between (66) and (67), 
since it is (66) which is isolated from the action. 
Writing
\eq
C^{-1}_{\ga}\,-\,C^{-1}_{ls}\,=\de C_{\ga}\,-\,P_l
\eqe
we obtain for $\de C_{\ga}$ 
the sum of terms~:
\eq
\de C_{\ga}\,=\,\sum_{i=1}^{i=4} \de C_i(\ga)\,\,,
\eqe
\[
\de C_1(\ga)\,=\,- \,P_s (\sqrt{1+\pi})P_{\ga}(\sqrt{1+\pi})P_s\,\,,
\]
\[
\de C_2(\ga)\,=\,P_l(\sqrt{1+\pi})(1-P_{\ga})(\sqrt{1+\pi})P_s
\,+\,P_s(\sqrt{1+\pi})(1-P_{\ga})(\sqrt{1+\pi})P_l
\]
\[
\,+\,P_l(\sqrt{1+\pi})(1-P_{\ga})(\sqrt{1+\pi})P_l\,\,,
\]
\[
\de C_3(\ga)\,=\,(1-P_{\La})(\sqrt{1+\pi})(1-P_{\ga})(\sqrt{1+\pi})P_{\La}
\,+\,P_{\La}(\sqrt{1+\pi})(1-P_{\ga})(\sqrt{1+\pi})(1-P_{\La})
\]
\[
\,+(1-P_{\La})(\sqrt{1+\pi})(1-P_{\ga})(\sqrt{1+\pi})(1-P_{\La})\,\,,
\]
\[
\de C_4(\ga)\,=\,\sqrt{1+\pi}\,\vep \,P_{\ga}\,\sqrt{1+\pi}\,\,.
\]
Having introduced the final covariance we may now rewrite the
expression for the two-point function based on the Gaussian measure
$d\mu_{\ga}$ with covariance $C_{\ga}$ normalized such that 
\eq
\int d\mu_{\ga}(\tau)\,=\,1\,.
\eqe
Since our covariance is configuration dependent there will be a
change of normalization of the functional integral when changing 
the l/s-assignment. Relative to the situation where $\ga\,=\,\emptyset$
this normalization factor is given by [13]
\eq
Z_{\ga}\,=\,{\det}^{1/2}(C_{\ga}/C_0)\,\,,
\eqe
where $C_0$ is given below (76).
Taking into account this factor we may rewrite (25) as
\eq
S^{\La}_2(x,y)\,=\,\frac{\sum_{l,s} Z_{\ga}\int d\mu_{\ga}(\tau)
\,(\frac{1}{p_{reg}^2+m^2+ig\tau \chi_{\La}})(x,y)\,G_{\ga}}
{\sum_{l,s}Z_{\ga}\int d\mu_{\ga}(\tau)\,\,G_{\ga}}\,,\quad \ga=\ga(l)\,.
\eqe
For the action $G_{\ga}$ we find collecting the results of the previous
manipulations:
\eq
G_{\ga}\,=\,\Pi_{\De \in \La_l} \,\theta_{\De}^l(\tau)\,
\Pi_{\De \in \La_s} \,\theta_{\De}^s(\tau)
\,\,
e ^{-1/2\int_{\La_l}\tau ^2}\,
{\det}_3^{-N/2}(1+iA_s)\,{\det}_2^{-N/2}(1+\frac{1}{1+iA_s}iA'')
\eqe
\[
\times e^{-R\int_{\bbbr^2-\La}\tau ^2}\,
\,\,e ^{1/2\,(\tau,\,\,\de C_{\ga}\,\tau)}\, \,.
\]
Note that we get indeed  
${\det}_3^{-N/2}(1+iA_s)\,{\det}_2^{-N/2}(1+\frac{1}{1+iA_s}iA'')$
after using the gap equation and absorbing the quadratic part of 
${\det}^{-N/2}(1+iA_s)$ since
\eq
TrA\,=\,TrA_s\,+\,TrA_l\,,\quad TrA_l\,=\,Tr(\frac{1}{1+iA_s}A_l)\,,
\eqe
on using $TrA'\,=\,0$ and $A_s\,A_l\,=\,0$. 

We first analyse the covariance $C_{\ga}$. Then we bound the
normalization factors $Z_{\ga}$ and the correction terms $\de C_{\ga}$.
Finally we bound the large field determinant.
Calling $C_0$ the covariance $C_{\ga}$ for the case that $\ga
\,=\,\emptyset$ which means
\eq
C_0\,=\,\frac{1}{\sqrt{1+\pi}}\,\frac{1}{1+f}\,\frac{1}{\sqrt{1+\pi}}
\eqe
we may write the inverse of (67) as
\eq
C_{\ga}\,=\,C_0\,+\,C_0(C_0^{-1}\,-\,C_{\ga}^{-1})C_{\ga}\,=\,
C_0\,\sum_{r=0}^{\infty}[(C_0^{-1}\,-\,C_{\ga}^{-1})C_0]^r
\eqe
\[
=\,
\frac{1}{\sqrt{1+\pi}}\,\frac{1}{\sqrt{1+f}}\sum_{r=0}^{\infty}
[\frac{1}{\sqrt{1+f}}\,P_{\ga}\,(1-\vep)\,\frac{1}{\sqrt{1+f}}]^r\,
\frac{1}{\sqrt{1+f}}\,\frac{1}{\sqrt{1+\pi}}
\]
\[
=\,C_0\,+\,
\frac{1}{\sqrt{1+\pi}}\,\frac{1}{1+f}\sum_{r=0}^{\infty}
P_{\ga}\,(1-\vep)[\frac{1}{1+f}\,P_{\ga}\,(1-\vep)]^r
\,\frac{1}{1+f}\frac{1}{\sqrt{1+\pi}}\,\,.
\]
The sums are  obviously norm-convergent. 
At this stage the support properties of $1/(1+f)$ (11) become very helpful.
They imply that in position space $C_{\ga}$ may be written 
in terms of a simple sum
over disconnected pieces with support restricted to (a neighbourhood of)
$\ga_i$. 
We obtain
\eq 
C_{\ga}\,=\,C_0\,+\,C^{\ga}\,,\quad C^{\ga}:=\,\sum_{i=1}^nC^{\ga_i}\,\,,
\eqe
\eq
C^{\ga_i}\,~:=\,
\frac{1}{\sqrt{1+\pi}}\,\frac{1}{1+f}\,
P_{\ga_i}\,(1-\vep)\sum_{r=0}^{\infty}
[\frac{1}{1+f}\,P_{\ga_i}\,(1-\vep)]^r
\,\frac{1}{1+f}\frac{1}{\sqrt{1+\pi}}\,\,.
\eqe
If we had only imposed exponential fall-off for  $1/(1+f)\,$, 
arbitrarily many terms coupling the
various $\ga_i$ would have appeared. They could be shown to be small
using the distance between the various $\ga_i$ of size $\sim M$ and
the fall-off of $1/(1+f)$ and of 
$\frac{1}{\sqrt{1+\pi}}$, but still they would be a
nuisance. The fall-off properties of $C_0$ have been analysed in Lemma
1. The complications stemming from nonempty $\ga$ are controlled
easily in\\
{\bf Lemma 3:} The kernel $C^{\ga}$ for $\ga \,\neq \, \emptyset$
satisfies the following estimates:\\
\eq
|C^{\ga}(x,y)|\,\leq\,O(1)\,N^{2/5}\exp \{-2m(dist(x,\ga)+dist(y,\ga))\}\,\,. 
\eqe
For $x,\,y \in \La -\Ga$ or $x\in \Ga_i\,,\,\,y \in\Ga_j$ with $i\neq j$
we find:
\eq
|C^{\ga}(x,y)|\,\leq\,O(1)\frac{1}{N^{18/5}}
\eqe
and for $x\in \Ga$, $y\in \La-\Ga$
\eq
|C^{\ga}(x,y)|\,\leq\,O(1)\frac{1}{N^{8/5}}\,\,.
\eqe
Finally we have
\eq
|C_0(x,y)|\,\leq\, O(1)\exp \{-2m|x-y|\}\,\,.
\eqe
{\sl Proof:} We have to control the contribution of 
the infinite sum over $r$ in (79). We abbreviate 
\eq
O \,=\, P_{\ga_i}\,(1-\vep)\sum_{r=0}^{\infty}
[\frac{1}{1+f}\,P_{\ga_i}\,(1-\vep)]^r\, ,\,\,
B\,=\, \frac{1}{\sqrt{1+\pi}}\,\frac{1}{1+f}
\eqe
so that 
\eq
C^{\ga_i}\,=\,B\,O\,B^*
\eqe
Obviously $||O||\,\leq\,N^{2/5}\,$ and $||B||\,\leq\,1$. Furthermore
the kernel of  $B$ is continuous and pointwise bounded by 
$O(1)$. By inserting characteristic functions 
of squares $\De$ between $B$ and $O$ and between $O$ and $B^*$, summing 
over the squares, using the fall-off properties of the kernels  and bounding
\eq
|(\chi_{\De},\,\,O\,\chi_{\De'})|\,\leq\, N^{2/5}
\eqe
we then arrive at the bounds stated in Lemma 3.
For the required properties of the kernels see Lemma 1 and  
(11). The minimal distances of points fulfilling the conditions
specified in Lemma 3 follow from the definitions (54)-(59). \qed\\
We remark that the bounds in Lemma 3 could be somewhat improved on
by using methods similar to those employed in the proof of Lemma 4.
We do not do so because this improvement 
would not strengthen our final bounds anyway.
Note in particular that the cluster expansion will be performed such
that only $C^{\ga}$-terms bridging the gap between $\ga$ and
$\La-\Ga\,$ will be produced.
Now we are going to bound the factors $Z_{\ga}$.\\[.1cm]
{\bf Lemma 4:} Let $|\ga|$ denote the volume of $\ga$. Then
\eq
\,\,1\,\leq\,Z_{\ga}\,\leq \,e ^{O(1)|\ga|}\,\,.
\eqe
{\sl Proof~:} Using (76),(78),(79) we have 
\eq
Z_{\ga}\,=\,{\det}^{1/2}(C_{\ga}/C_0)\,=\,
{\det}^{1/2}(1+\frac{1}{C_0}\sum_i C^{\ga_i})
=\,{\det}^{1/2}(1+\sum_{i,r_i\geq 1}
[(1-\vep)\,P_{\ga_i}\,\frac{1}{1+f}]^{r_i})
\eqe
\[
=\,\Pi_i \,{\det}^{1/2}(1+\sum_{r_i\geq 1}
[(1-\vep)\,P_{\ga_i}\,\frac{1}{1+f}]^{r_i})
=\,\Pi_i\,{\det}^{1/2}(C_{\ga_i}/C_0)\,=\,\Pi_i \,Z_{\ga_i}\,\,.
\]
Again we used the support properties  of $1/(1+f)$ to factorize the
determinant. For $Z_{\ga_i}$ we now find
\eq
Z_{\ga_i}\,=\,
{\det}^{1/2}(1+\frac{(1-\vep)\,P_{\ga_i}\,\frac{1}{1+f}}
{1\,-\,(1-\vep)\,P_{\ga_i}\,\frac{1}{1+f}})
\,=\,
{\det}^{-1/2}(1-(1-\vep)\,P_{\ga_i}\,\frac{1}{1+f})
\eqe
\[
=\,\exp Tr\{(-1/2)\ln (1-(1-\vep)\,P_{\ga_i}\,\frac{1}{1+f})\}
\,=\,
\exp Tr \{ 1/2\sum_{r\geq 1}\frac{1}{r} (1-\vep)^r\,
[P_{\ga_i}\,\frac{1}{1+f}]^r\}\,\,.
\]
This expression implies $Z_{\ga}\,\geq \,1$.
On the other hand we may use Lemma 3' from [1], which says that for
an Hermitian trace class operator $A$ and orthogonal Projector $P$
we have the inequality~:
\eq
Tr(P\,A\,P)^r \,\leq \,TrP\, A ^r\,P\,\,.
\eqe
Applying this to $\frac{1}{1+f}$ and $P_{\ga_i}$ we may bound
\eq
Tr(\frac{1}{1+f}\,P_{\ga_i})^r \,=\,
Tr(P_{\ga_i}\,\frac{1}{1+f}\,P_{\ga_i})^r 
\eqe
\[
\leq\,
Tr(P_{\ga_i}\,(\frac{1}{1+f})^r\,P_{\ga_i})\,\leq\,
O(1)\,|\ga_i|\,\,.
\]
using the fact that\footnote{Unfortunately the factor of $r^{-1/2}$
appearing in (92) is falsely written as $2^{-r}$ in [1]. This mistake
fortunately is of no consequence however.}
\eq
\int d^2p\,(\frac{1}{1+f(p)})^r\,\leq\,O(1)\,
\int d^2p\,(\frac{1}{1+(p^2)^2})^r
\,\leq\, O(1) r^{-1/2}\,\,.
\eqe
Using this we obtain
\eq
Z_{\ga}\,\
\leq\,
\exp\{O(1)\,|\ga|\, \sum_{r\geq 1}\frac{1}{r^{3/2}} (1-\vep)^r\}
\,\leq\,\exp(O(1)|\ga|)\,\,.
\eqe
This proves Lemma 3. \qed \\ 
We now come to the bounds
on the correction terms $\de C_i(\ga)$ from (70).\\
{\bf Lemma 5:}\\
i) \,\,$\,\de C_1(\ga) \,\leq \, 0\,$  (as an operator),\\
ii) $\,\,\de C_3(\ga) \,\leq \,1\,+\,\pi(0)\,$, 
$\,\,\de C_4(\ga) \,\leq \,O(1) \,N^{-2/5}$ (as operators),\\
iii) 
$||\de C_2(\ga)|| \,\leq \,O(1)\,N^{-2}$,
$\,\,|\de C_2(\ga)(x-y)| \,\leq \,O(1)\, \inf\{e^{-2m|x-y|},\,N^{-2}\}
\,\,$.\\[.1cm]
{\sl Proof~:}
i) is immediately obvious from the positivity of $\pi$.\\
ii) The first statement is obvious since 
\eq
\de C_3(\ga)\,=\,   
(\sqrt{1+\pi})(1-P_{\ga})(\sqrt{1+\pi})\,-\,
P_{\La}(\sqrt{1+\pi})(1-P_{\ga})(\sqrt{1+\pi})P_{\La}\,\,.
\eqe
Note that $\de C_3(\ga)$ only enters through interactions with field
configurations of support outside $\La$, which will be suppressed anyway
when taking  $R\to \infty$,  (Prop.8,(114)).
The bound on $\de C_4(\ga)$ follows from the definition of $\vep$
in (68).\\
iii) The first statement in iii) follows from the exponential
fall-off of $\sqrt{1+\pi}\,$ (Lemma 1) and the fact that
$dist(\La_l,\,(\bbbr^2-\ga))\,\geq \,{\ln N \over m}$ (see (58)).
This implies a bound on $\de C_2(\ga)$ of the form in iii), closer
inspection shows that O(1) is basically given by $m^{-3}$, two powers
coming from the integration over the kernel bridging the distance gap
and one coming from a norm bound on the second $\sqrt{1+\pi}\,$. 
The second statement in iii) also 
follows from the definition of $\ga$ 
and from the fall-off of the kernel of $\sqrt{1+\pi}\,$.
\qed \\
Now we come to the bound on the nondiagonal term
${\det}_2^{-N/2}(1+\frac{1}{1+iA_s}iA'')\,$ in the action (74). 
We need to get a suitable bound for this term which is sufficiently
stable under the modifications caused by the cluster expansion parameters. 
We (temporarily) introduce the operator $B$ through
\eq
B\,=\,\frac{1}{1+iA_s}iA''\,=\,\frac{1}{(1+iA_s)(1-iA^*_s)}(i+A^*_s)A''\,\,.
\eqe
Using the facts that the $A$-operators have real expectation values in
real Hilbert space, that $TrA_s^nA''\,=0$ and cyclicity we find
\eq  
|{\det}_2^{-1}(1+B)|\,=\,|{\det}^{-1}(1+B)|\,=\,
|{\det}^{-1}(1+B^*)|\,=\,
{\det}^{-1/2}(1+D)\,,
\eqe
where
\eq
D\,=\,B\,+\,B^*\,+\,B^* B\,\,.
\eqe
Now we may apply the norm bound on $A_s$ from Proposition 2 to realize
that $B$ coincides with $iA''$ up to small corrections, more
precisely:\\[.1cm]
{\bf Lemma 6:} For $\vp \,\in\, {\cal L}^2(\La)$ we find
\eq
B\vp\,=\,iA''\vp\,+\,\de \,A''\vp\,,\quad 
B^*\vp\,=\,-iA''^*\vp\,+\,A''^*\de^*\vp\,\,, 
\eqe
where the operator $\de$ is bounded in norm as
\eq
||\de|| \,\leq\,(1+\alpha)||A_s||\,\leq\, (1\,+\,\al)\,N^{-2/5}\,<\!<1
\eqe
with suitable $0 < \alpha <\!<1$.\\[.1cm]
{\sl Proof:}
Since we have
\eq
\de\,=\,i(\frac{1}{1+iA_s}\,-\,1)
\eqe
the statements of the Lemma follow directly from Proposition 2,
where $\alpha$ may be chosen  to obey an upper bound 
of size $\sim ||A_s||$.\qed 

Now we can also bound the operator $D$. For  $\vp \,\in\, {\cal
L}^2(\La)$ normalized to one
we
obtain 
\eq
(\vp,\,D\vp)\,=\,
i(\vp,\,A''\vp)\,-\,i(A''\vp,\vp)\,+\,
(\vp,\,\de\,A''\vp)\,+\,(\de\,A''\vp,\,\vp)
\eqe
\[
\,+\,(A''\vp,\,A''\vp)\,+\,
(A''\vp,\,(\frac{1}{(1-iA^*_s)(1+iA_s)}\,-\,1)A''\vp)\,\,.
\]
Since the first two terms drop out, this entails
\eq
(\vp,\,D\vp)\,\geq\,(1-\eta)
||A''\vp||^2_2\,-\,\eta ||A''\vp||_2\,\geq\,
-\frac{4}{25}\frac{\eta^2}{1-\eta}
\eqe
with the choice 
\eq
\eta\,=\,2\,||\de||\,<\!< 1\,\,.
\eqe
Splitting the selfadjoint operator $D$ into its negative part
$D_-$ and its nonnegative part $D_+$
\eq
D\,=\,D_+\,-\,D_-
\eqe
we thus have obtained that 
\eq
0\,\leq \,D_-
\,\leq\,\frac{4}{25}\frac{\eta^2}{1-\eta}\,\leq\, N^{-4/5}\,\,.
\eqe
Using this we may now proceed to a bound on
$|{\det}_2^{-1}(1+B)|\,=\,{\det}^{-1/2}(1+D)$.
We find
\eq
{\det}^{-1/2}(1+D)\,= \,
e^{-1/2TrD}\,{\det}_2^{-1/2}(1+D)
\eqe
\[
\,=\,
e^{-1/2TrB^*B}\,\,{\det}_2^{-1/2}(1+D_+)\,\,{\det}_2^{-1/2}(1-D_-)
\,\leq\,
{\det}_2^{-1/2}(1-D_-)\,\,.
\]
Here we used again the fact that $TrB\,=\,i\,TrA_l\,=\,-TrB^*$.
Evaluating the trace of $D_-$ in an eigenbasis of $D_-$ one may easily
establish the bound 
\eq
{\det}_2^{-1/2}(1-D_-)\,\leq\,
\exp(1/2Tr(\frac{D_-^2}{2-||D_-||-D_-}))
\,\leq\,
\exp (\frac{1}{4-4||D_-||} TrD_-^2)\,\,.
\eqe
To verify the first inequality one observes that for
 $x>\!-\vep>\!-1$ we have $x-\ln(1+x)\leq\frac{x^2}{2-\vep+x}\,$.
Now it remains to bound  $Tr D_-^2\,$. We call $\{\vp_-\}$ a suitable
set of normalized eigenfunctions of $D_-$ and  find
\eq
TrD_-^2\,=\,\sum_{\vp_-}(\vp_-,\,D^2\vp_-)\,=\,
\sum_{\vp_-}[(\vp_-,\,D\vp_-)]^2
\eqe
\[
\leq\,
\sum_{\vp_-}(\frac{4}{5}\,\eta\,||A''\vp_-||)^2
\,\leq\,
\frac{16}{25}\,\eta^2 \,Tr(A''^*A'')\,\leq\,
\frac{16}{25}\,\eta^2 \,Tr(A^*A)\,
\]
\[
\leq\,\frac{16}{25}\,\eta^2 \,g^2(\int_{\La}\tau^2)(\int\frac{d^2p}{(2\pi)^2}
(\frac{1}{p^2_{reg}+m^2})^2)\,\leq\,
O(1)N^{-4/5}\,g^2\,(\int_{\La_s}\tau^2\,+\,\int_{\La_l}\tau^2)\,\,.
\]
In the first inequality in (108) we made use of (102).
It is admittedly pedantic to insist on factors as $(4/5)^2$ in our
context. To pass from  
$Tr(A''^*A'')\,$ to the expression in the last line in (108) it is
sufficient to take away the projectors $P_l\,$ or $P_s\,$ and thus to
bound $Tr(A''^*A'')\,$ in terms of $Tr(A^*A)\,$ which is given by the
double integral. We have obtained \\[.1cm] 
{\bf Lemma 7:}
\eq
|{\det}_2^{-N/2}(1+\frac{1}{1+iA_s}iA'')|\,\leq\,
e ^{O(1)N^{-4/5}\,\int_{\La}\tau^2\,}\,\,.
\eqe

Finally we may also bound somewhat further the terms
$(\tau, \,\de C_3(\ga)\tau)$ and $(\tau, \,\de C_4(\ga)\tau)$
appearing in $G_{\ga}$. Writing $\hat{\tau}:=\,\tau\,\chi_{\bbbr^2-\La}$
we have
\eq
1/2\,(\tau, \,\,\de C_3(\ga)\,\tau)\,=\,
(\hat{\tau},\,\,\de C_3(\ga)\,\tau)\,+\,
1/2\,(\hat{\tau},\,\,\de C_3(\ga)\,\hat{\tau})\,\,.
\eqe
We may then bound 
\eq
(\hat{\tau},\,\,\de C_3(\ga)\,\tau)\,+\,
1/2\,(\hat{\tau},\,\,\de C_3(\ga)\,\hat{\tau})\,\leq
R/2 \;(\hat{\tau},\,\,\hat{\tau})
\,+\,\frac{O(1)}{R}\,(\tau \,\chi_{\La},\,\,\tau \,\chi_{\La})
\eqe
for $R$ large enough, using Lemma 5. As for $ \de C_4(\ga)$ we
find 
\eq
(\tau,\,\,\de C_4(\ga)\,\tau)\,\leq\,
O(1)N^{-2/5}\,\bigl[ (\tau \,\chi_{\La},\,\,\tau\,\chi_{\La}) 
\,+\,(\hat{\tau},\,\,\hat{\tau})\bigr]\,\,.
\eqe
Now we dispose of a complete control of the action $G_{\ga}$ from (74)
and may collect our in findings in\\[.1cm] 
{\bf Proposition 8:} For $R$ large enough we have
\eq
Z_{\ga}\,\,G_{\ga}(\tau)\,\leq\,
e ^{-{49 \over 100}\,\int_{\La_l}\tau ^2}\,
e ^{O(1)N^{-2/5}\int_{\La_s}\tau ^2}\,
\, e^{-R/2\,\int_{\bbbr^2-\La}\tau ^2}\,\,.
\eqe
Thus we now take the limit $R\to\infty$ and absorb 
the term $e^{-R/2\,\int_{\bbbr^2-\La}\tau ^2}$ in the covariance, 
which implies that we may replace
\eq
C_{\ga}\,\to \,P_{\La}\,C_{\ga}P_{\La}
\eqe
and restrict the action to configurations $\tau(x)$ with $supp\,\tau
\subset \La\,$.\\[.1cm]  
{\sl Proof:} 
The proof concerning (114) is to be found e.g. in [13], so we have only to   
gather  the pieces for the proof of (113). In the first term
$e ^{-{49 \over 100}\,\int_{\La_l}\tau ^2}\,$ we collected together 
the contribution
$e ^{-1/2\,\int_{\La_l}\tau ^2}\,$ from (74)
the $\La_l$-contribution from (109), the term
from the bound on $Z_{\ga}$ in Lemma 4, where we used
\eq
O(1) \,|\ga|\,\leq\, O(1)\,({\ln N \over m})^2\,|\La_l|\,\leq\,
O(1)\,({\ln N \over  m})^2\,N^{-1/6}\,\int_{\La_l}\tau ^2\,\,,
\eqe
and the contributions from (111) and (112) in $\La_l\,$. 
Finally we also absorbed in this term a
contribution coming from $\de C_2(\ga)$.
In the term $e ^{O(1)N^{-2/5}\int_{\La_s}\tau ^2}\,$ we have
absorbed the contribution in $\La_s$ from (109),(111),(112) and 
again a contribution coming from $\de C_2(\ga)$. Finally we absorbed 
the one from the
bound on $|{\det}_3^{-N/2}(1+iA_s)|$ to be derived now: \\
$|{\det}_3^{-N/2}(1+iA_s)|$ can be bounded using the inequality
\eq
|Tr A ^n|\,\leq\, ||A ^{n-2}||\,Tr(A ^*A)
\eqe
valid for any traceclass operator $A$ and $n\,\geq\, 2$.
To bound  $||A_s^{n-2}||$ we use  Proposition 2. 
We may restrict to
$n=3$, the subsequent terms in the expansion of ${\det}_3^{-N/2}(1+iA_s)$
being much smaller
\eq
|TrA_s^3|\,\leq \, ||A_s||\,Tr(A_s^*A_s)\,\leq
N^{-2/5}N^{-1}\,O(1)(\tau \chi_s,\,\pi(0)\,\tau \chi_s)\,\leq\,
O(1) N^{-7/5}\,\int_{\La_s}\tau ^2\,\,.
\eqe 
With the help of the previous remarks and this relation 
we can verify  the bound \\
$|{\det}_3^{-N/2}(1+iA_s)|$ $\leq$
$e ^{O(1)N^{-2/5}\int_{\La_s}\tau ^2}\,$.
 \qed
The following result now is immediate.\\[.1cm]
{\bf Corollary:} Reducing the volume $\La$ to a single square $\De$ 
equipped with a small field condition (42) we find 
\eq
Z^{\De}\,=\, \int \, d\mu^{\De}(\tau) G^{\De}\,
\,=\,1\,+\,o(N^{-1/5})\,\,,
\eqe
where $G^{\De}$ is the integrand from (26)
restricted to the single small field square volume, 
and $ d\mu^{\De}(\tau)$ is the normalized measure 
with covariance $\chi_{\De}\,C_0\,\chi_{\De}$.\\[.1cm]
The statement follows from the bound (113) restricted to one small
field square in $\La$.\\[.1cm]
It will be useful later on to bound  the large field contribution 
in (113), r.h.s. by a product of suppression factors
in probability per square $\De \in \Ga$. 
If (43) holds we may write 
\eq
\int_{\De\in \La_l} \tau ^2 \,\geq\,
1/2\int_{\De\in \La_l} \tau ^2\,+\,{3 \la K \over 8} N^{1/6}
\,\,\mbox{ and }\,\,
\int_{\De\in \La_{l^n}} \tau ^2 \,\geq\,
1/2\int_{\De\in \La_{l^n}} \tau ^2\,+\,{3 \la K \over 8} N^{n/6}\,\,.
\eqe
Therefore we obtain\\[.1cm]
{\bf Lemma 9:} 
\eq 
e ^{-{49 \over 100}\,\int_{\La_l}\tau ^2}\,\leq\,
e ^{-1/4\,\int_{\La_l}\tau ^2}\,\, 
e ^{-N^{1/8}|\Ga ^e|}\,\,\prod_{\De \in \La_{l^n}} 
e ^{-N^{{n-1 \over 8}}}\,\, .
\eqe
{\sl Proof:} It suffices to observe that for $N$ sufficiently large
we have (using (42),(45)) 
\[
O(1) ({n\ln N \over m })^{-2} N^{n/6}\,\geq\,N^{{ n \over 8}}\,\,.
\]
\qed
Now we have sufficient control of the action to start with the expansions.

\section{The Expansions, Proof of Mass Generation}
\subsection{The General form of the Expansions} 
The cluster expansion
allows to control the spatial correlations of the model. When combined
with a subsequent Mayer expansion, which frees the clusters 
from their hard core constraints, it allows to
take the thermodynamic limit and to bound
the decay of the correlation  functions. We proceed similarly as in
[1] and use the general formalism for cluster expansions presented in [18], 
which in turn is an  elaboration on a theme which has been the subject  
of several seminal papers by Brydges and collaborators over more than
a decade. We apply in particular the Brydges-Kennedy formulae [15].
For general references on cluster expansions see also [13], where  
the presentation is close to the original way of introducing cluster
expansions in constructive field theory, and [14], [19], 
which are close to our way of presentation.

The cluster expansion is a technique to select explicit connections
between different spatial regions. The best formulas for the clusters involve
trees, which are the minimal way to connect abstract objects together.
We call the subsequent formulae forest formulae, the  forests 
generally consisting of several disconnected trees.
The basic building blocks of our expansion are the large
field blocks $\Ga^e_i$ (62) composed of (generally many) 
large field squares and their
security belts, and the individual small field squares $\De$ from
\eq
S:=\La\,-\,\Ga ^e\,\,.
\eqe
From the point of view of the presentation it seems
advantageous to connect together these large field blocks $\Ga ^e_i$ 
already by a first cluster expansion, and then to proceed to a second
one, the building blocks of which are given by the outcome of the first.
Then the expansion really connects together unit size squares which
allows to somewhat unify the language as regards convergence
criteria etc. In view of the existence of the excellent presentations
to be found in [14]-[19] and since we stick very closely to [18]
we hardly give indications on the proofs of cluster expansion
formulae here.
The general forest formula 
we are going to use will be given now. We introduce the
following notation:\\ 
Let $I$ be a finite index set (in our context the set $\La$ of the 
squares $\De \in \La$) and $P(I)$ the set of all
unordered pairs $(i,j)\in I\times I$, $i\ne j$. A (unordered) forest $\cal F$
on $I$ is a subset of $P(I)$ which does not contain loops $(i_{1},\,i_{2})
\, \ldots \,(i_{n},i_{1})$. Any such forest splits as a single union 
of disjoint 
trees, and it gives also a decomposition of $I$ into $|I|-|\cal F|$
clusters (some of them possibly singletons). The 
non-trivial clusters are connected
by the (non-empty) trees of the forest.
Let $H$ be a function of variables $x_{ij}$, $ij \in P$.
Then the following forest formula due to Brydges is proven in [18]:

\eq
H(1,...,1) \,= \,
\sum_{\cal F} \bigl(\prod _{l\in \cal F}\int_{0}^{1} dh_{l}\bigr)
\biggl( \bigl(\prod _{l\in \cal F}  {d \over dx_{l}}\bigr) H
\biggr) ( h^{\cal F}_{ij} (h))\,\,,
\eqe
where 
\eq
h^{\cal F}_{ij} (h) \,=\, {\rm inf} \{ h_{l} , l \in L_{\cal F}(i,j) \}
\eqe
and $L_{\cal F}(i,j)$ is the unique path in the forest $\cal F$ connecting $i$
to $j$. If no such path exists, by convention $h^{\cal F}_{ij} (h) =0$.

This interpolation formula will subsequently be applied to our
expression for the two-point function,
more precisely to the summands in the numerator
and denominator of (73) with given $l/s$-assignments.
As mentioned we proceed in two
steps. The first rather trivial one is to connect together 
the squares in the components $\Ga ^e_a$ of $\Ga ^e$.  
Let $P_{\Ga ^e}$
be the set of all pairs $(i,j)$ of distinct squares in $\Ga ^e$. We define
$\vep_{ij}=0$, if $\De_i \cap \De_j = \emptyset$ or  
if $\De_{i}$ and $\De_{j}$ belong to different components 
$\Ga_{a(i)}^e\,\neq\,\Ga_{a(j)}^e$ of $\Ga ^e$,
$\vep_{ij}=1$ otherwise, and $\eta_{ij} =1 -\vep_{ij}$. 
Our first forest formula is simply 
\eq
1 \,=\, \sum_{{\cal F}_{1}} \prod_{l \in {\cal F}_{1}} 
\bigl( \vep_{l}\int_{0}^{1} dh_{l} \bigr)
\prod _{l \not\in {\cal F}_{1}}  (\eta_{l} + \vep_{l} h^{{\cal F}_1}_{l} (h))
\,\,.\eqe
This follows directly from the application 
 the forest formula to $H(\{x_{ij}\}) = 
\prod_{ij \in P_{\Ga ^e}} (x_{ij}\vep_{ij}+\eta_{ij})$ 
using that here $H(1,...,1)=1$. 

The only non-zero terms in this formula are those for which the
clusters associated to the forest ${\cal F}_{1}$ are exactly the set
of connected components $\Ga^e_a$ of the large field region.
Indeed they cannot be larger because of the factor 
$\prod_{l \in {\cal F}_{1}}  \vep_{l}$, nor can they be smaller
because of the factor  
$\prod _{l \not\in {\cal F}_{1}}  
(\eta_{l} + \vep_{l} h^{{\cal F}_1}_{l} 
(h)) $,
which is zero if there are some neighbours belonging to the same component
(for which $\eta_{ij}=0$)
belonging to different clusters (for which $h^{{\cal F}_1}_{ij}
(h)=0$). Therefore this formula simply associates connecting trees of 
``neighbour links'' to each such connected component, but in a
symmetric  way  without arbitrary choices. We remark finally
that in (124) the interpolated factors
$\prod _{l \not\in {\cal F}_{1}}  (\eta_{l} + \vep_{l} h^{{\cal
F}}_{l} (h))$  after giving the necessary 
constraints on
the clusters can be bounded simply by 1.

The second cluster expansion links together the previous clusters
by interpolating all the non-local kernels in the theory. It
gives a forest formula which is an extension of the first one.
We consider all non-local kernels in our theory, that is 
\eq
{1 \over \sqrt {1 + \pi}} \, \, , \quad 
\sqrt {1 + \pi}\,\,, \quad   {1 \over  {p^2 + m^2}}\,\,\,.
\eqe
Note that due to the support property (11) of 
$1/(1+f)$ and our choice of treating each $\Ga ^e_a$
as one connected block of the
second expansion, we need not interpolate 
$C_{\ga} \,= \,C_{0} \,+\, C^{\ga}$ as a whole: When all kernels
appearing in (125) are interpolated such that they do not connect
any more different clusters of the second expansion,
then $C_{\ga}$ - with these interpolated kernels replacing the
noninterpolated ones in the expression for $C_{\ga}$ 
- does not connect different clusters either.
The three kernels from (125) will be generically called $K$.
Now the second expansion takes into account the 
connections built by the
first, i.e. it interpolates  only the links 
\eq
K_{l}(x,y)\,=\, 
K_{ij}(x,y) \,=\,\De_{i}(x)\, K(x,y)\, \De_{j}(y)
\eqe
for squares which belong to {\it different} clusters of the 
first forest.
Let $Z(K, \Ga ^e,\La)$ be a generic name for the quantities we want to
compute, namely the numerator and denominator in (73). 
Then the second forest formula gives:
\eq
Z(K, \Ga ^e,\La) \,=\, 
\sum_{{\cal F}_{1}} \prod_{l \in {\cal F}_{1}} 
\bigl( \vep_{l}\int_{0}^{1} dh_{l} \bigr)
\prod _{l \not\in {\cal F}_{1}}  (\eta_{l} + \vep_{l} h^{{\cal F}}_{l} (h))
\,\times 
\eqe
\[
\times\,\sum_{{\cal F}_{2} \supset {\cal F}_{1}} 
\prod_{l \in {\cal F}_{2}-{\cal F}_{1}} \bigl( \int_{0}^{1} dh_{l} \bigr) 
\prod_{l \in {\cal F}_{2}-{\cal F}_{1}}
\bigl({d \over dx_{l}} \bigr) 
Z (K(\{h_{{\cal F}_{2}-{\cal F}_{1}}\}),\La)\,\,,  
\]
where $ Z (K(\{h_{{\cal F}_{2}-{\cal F}_{1}}\}),\La)$ is a functional integral
with interpolated kernels $K(\{h_{{\cal F}_{2}-{\cal F}_{1}}\})$. These
interpolated kernels are defined by
$K(\{h_{{\cal F}_{2}-{\cal F}_{1}}\})\,=\,
h^{{\cal F}_{1},{\cal F}_{2}}_{l} (h) K_{l}(x,y)$,
where $h^{{\cal F}_{1},{\cal F}_{2}}_{l} (h)$ 
is the $\inf$ of the $h$ parameters of the lines
of ${\cal F}_{2}-{\cal F}_{1}$ on the unique path in ${\cal F}_{2}$ 
joining $\De_{i}$
to $\De_{j}$ (if $l= (i,j)$). 
Again if no such path exists, by convention 
$h^{{\cal F}_{1}, {\cal F}_{2}}_{l}(h)\,=\,0$.
In other words the path is computed with the full forest, but 
only the parameters of the forest ${\cal F}_{2}-{\cal F}_{1}$ 
are taken into account
for the interpolated non-local kernels.

The product $\prod_{l \in {\cal F}_{2}-{\cal F}_{1}}
\bigl( {d \over dx_{l}}  \bigr)$ is a short notation for an operator which
derives with respect to a parameter $x_{l}$ multiplying 
$K_{l}$ where $K$ is any of the non-local kernels,
and then takes $x_{l}$ to 1. 
Therefore the action of $\prod_{l \in{\cal F}_{2}-{\cal F}_{1}}
\bigl( {d \over dx_{l}}  \bigr)$ creates
the product $\prod_{l \in {\cal F}_{2}-{\cal F}_{1}} K_l$
(with summation over the finite set of possible $K$'s),
multiplied either by functional derivatives hooked to both ends
(for the case where the derivatives apply to the measure and are
evaluated by partial integration) 
or by other terms descended from action exponential, if the derivatives
apply directly to the action. In 
section IV.4 we give the list of the corresponding derived ``vertices''
produced by these derivatives.
The important fact to be shown is that because these derivatives
act on terms which a carry a factor $N^{-x}\,,\,\,x>0\,$, in fact to each
such vertex, hence to each link of this second expansion, is associated
a factor which tends to zero as $N\to \infty$.

It is an important property of the forest formulas of this type that they
{\it preserve positivity} properties [18], so that
if $K$ is a positive operator, $K(\{h_{{\cal F}_{2}-{\cal F}_{1}}\})$ 
is also positive.
This is not obvious at first sight form the infimum rule of (123), 
but it is true because
for any ordering of the $h$ parameters (say $h_{1}\le ...\le h_{n}$) 
there is a way ({\it which varies
with the ordering}) to rewrite the interpolated $K(h)$ as an explicit sum of
positive operators [18]: 
\eq
K(h)= \sum_{p}(h_{p}-h_{p-1}) \sum_{q=1}^p\chi_{p,q}K\chi_{p,q} 
\eqe
The functions $\chi_{p,q}$ are  
the characteristic functions of the clusters built with the part of the
forest made of lines $p$, $p+1$,...,$n$. For us 
(as for anyone interpolating Gaussian measures) this
preservation of positivity is crucial when the covariance $C_{\ga}$
is interpolated.

\subsection{The Cluster amplitudes. Factorization}
From (127) we realize that the quantities $Z(K,\Ga ^e,\La)$
factorize over contributions, the mutually disjoint 
supports of which - to be called
polymers - are the blocks connected together by the links of the
disjoint trees in the forest $\,{\cal F}_2$. So they
take the form 
\eq
A(K, \Ga ^e,Y) \,=\, 
\sum_{\mbox{trees }\{{\cal T}_{1a}\}=:{\cal T}_1}
\,\, \prod_{l \in {\cal T}_{1}} 
\bigl( \vep_{l}\int_{0}^{1} dh_{l} \bigr)
\prod _{l \not\in {\cal T}_{1}}  (\eta_{l} + \vep_{l} h^{{\cal T}_1}_{l} (h))
\,\times 
\eqe
\[
\times\,\sum_{\mbox{trees }{\cal T}_{2} \mbox{ on } Y,\,{\cal T}_{2}
\supset {\cal T}_{1}}\,\,\,\, 
\prod_{l \in {\cal T}_{2}-{\cal T}_{1}} \bigl( \int_{0}^{1} dh_{l} \bigr) 
\prod_{l \in {\cal T}_{2}-{\cal T}_{1}}
\bigl({d \over dx_{l}} \bigr) A (K(\{h_{{\cal T}_{2}-{\cal T}_{1}}\}),Y)\,\,,  
\]
The trees ${\cal T}_{1a}$ join together the  
connected subsets of $Y\cap
\Ga ^e_a$, their union, called ${\cal T}_1$, (which in fact is a forest)  
becomes a subset of a single tree when adding the links from 
${\cal T}_{2}-{\cal T}_{1}$. The trees ${\cal T}_2$ connect together 
all of the polymer $Y$, so they have $|Y|-1$
elements. Then (similarly as above (122))
  $A(K(\{h_{{\cal T}_{2}-{\cal T}_{1}}\}),Y)$ 
is a functional integral
with interpolated kernels $K(\{h_{{\cal T}_{2}-{\cal T}_{1}}\})$. These
kernels are defined by 
$K(\{h_{{\cal T}_{2}-{\cal T}_{1}}\})\,=\,
h^{{\cal T}_{1},{\cal T}_{2}}_{l} (h) K_{l}(x,y)$,
where $h^{{\cal T}_{1},{\cal T}_{2}}_{l} (h)$ 
is the $\inf$ of the $h$ parameters of the lines
of ${\cal T}_{2}-{\cal T}_{1}$ on the unique path in ${\cal T}_{2}$ 
joining $\De_{i}$ to $\De_{j}$ for $l= (i,j)$. 

Now regarding more explicitly  the two-point function (73)
we get the following formula  as result of the cluster expansion:
\eq
S^{\La}_{2}(x,\,y) \,=\,
 { \sum_{l} \prod_{a} Z_{\ga_{a}} 
\sum\limits_{q, Y_{i}^{l} \atop
Y_{i }\cap Y_{j} =0, \cup_{i}Y_{i}=\La} 
A^{l}(Y_{1}, x,\,y)
(1/(q-1)!)\prod_{i=2}^{q}  A^{l}(Y_{i})
\over  \sum_{l} \prod_{a} Z_{\ga_{a}} 
\sum\limits_{q, Y_{i}^{l} \atop
Y_{i }\cap Y_{j} =0, \cup_{i}Y_{i}=\La}
(1/q!)\prod_{i=1}^{q}  A^{l}(Y_{i}) }
\eqe
with the following explanations~:\\
1) The amplitudes for the polymers depend on the choice $l$ of the large
field region. By shorthand notation $l$ stands for the infinite series
of possible choices $\,s,\,l^1,\,l^2,\ldots\,$. Correspondingly the 
sum $\sum_l$ stands for the infinite sum over those choices.
We note already that there is no convergence problem associated with
this infinite sum due to the suppression factors (120).\\
2) The difference between the numerator and the denominator in (130)
is that in the numerator there is {\it one} external polymer depending
on the source points  $x$ and $y$. 
Note that there is no nonzero contribution in which the points $x$ and $y$ lie
in two distinct polymers. This would necessitate to cut the factor
$(\frac{1}{p^2+m^2+ig\tau  \chi_{\La}})(x,y)\,$ into a product
of two pieces of disjoint support\footnote{We write
\[
\frac{1}{p^2+m^2+ig\tau  }\,=\,
\frac{1}{1\,+\,\frac{1}{p^2+m^2}ig\tau}\,\frac{1}{p^2+m^2}
\]
and interpolate the kernel $1/(p^2+m^2)$, see also 
the proof of the Theorem below .}, one containing $x$ and the other
$y$. Such a contribution obviously vanishes. 
The absence of such a contribution can be traced back to the symmetry
$\phi\,\to -\phi$ of the action (1).

Since by the rule of our cluster expansion,
each component $\ga_{a}$
of the large field region is contained
in exactly one polymer $Y$, we may absorb each normalization factor
$Z_{\ga_{a}}$ into its cluster, defining
\eq
\widetilde A(Y): = A(Y) \prod_{a/ \ga_{a} \subset Y} Z_{\ga_{a}}\,\,.
\eqe
The simplest cluster is a single small
field square $\De \subset S=\La-\Ga ^e$.\footnote{We assume 
the square not to contain
the external points $x,\,y$ which may be thought to lie far apart.}
 Due to (118) we find in this case
\eq
A_{0}(\De ) \,  = \,1 + o (N^{-1/5})\,\,.
\eqe
Therefore it is convenient to cancel out the background of
trivial single square small field clusters, hence to introduce
for a polymer $Y$ the normalized amplitude
\eq
a(Y) = {\widetilde A(Y)   \over \prod _{\De \subset Y} A_{0}(\De)}\,\,.
\eqe
Then we obtain the usual dilute polymer representation:
\eq
S_{2}(x,\,y) = { \sum_{l} 
\sum\limits_{q, Y_{i}^{l} \atop
Y_{i }\cap Y_{j} =0 } a^{l}(Y_{1}, x,\,y)
{1 \over (q-1)!}\prod_{i=2}^{q}  a^{l}(Y_{i})
\over  \sum_{l}  
\sum\limits_{q, Y_{i}^{l} \atop
Y_{i }\cap Y_{j} =0 }
(1/q!)\prod_{i=1}^{q}  a^{l}(Y_{i}) }\,\,.
\eqe
To get factorization we must analyze how the choice of $l$ affects
the cluster amplitudes. 
The choice of  the large field regions $\La_{l^n}$ 
for fixed $n$ is a local one, which means that
the constraints implied by the choice are of finite range. 
The sum over these choices therefore  can be 
absorbed into the value of (redefined) factorized amplitudes. 
Indeed we can replace the global sums over $s,\,l^1,\,l^2,\ldots\,$ 
by local ones:
\eq
\sum_{l} 
\sum_{q, Y_{i}^{l} \atop
Y_{i }\cap Y_{j} =0 } a^{l}(Y_{1}, x,\,y)
{1 \over (q-1)!}\prod_{i=2}^{q}  a^l(Y_{i})
\, = \,
\sum_{q, Y_{i}  \atop
Y_{i }\cap Y_{j} =0 } b(Y_{1}, x,\,y)
{1 \over (q-1)!}\prod_{i=2}^{q}  b (Y_{i})\,\,,
\eqe
\eq
\sum_{l} 
\sum_{q, Y_{i}^{l} \atop
Y_{i }\cap Y_{j} =0 } 
(1/q!)\prod_{i=1}^{q}  a^{l}(Y_{i}) = 
\sum_{q, Y_{i}  \atop
Y_{i }\cap Y_{j} =0 } 
(1/q!)\prod_{i=1}^{q}  b (Y_{i})
\eqe
with the explanations:\\
(i) The right sum is over all sets $\{ Y_{1}, ..., Y_{q}\}$
where the $Y_{i} $ are sets of $\De$'s, a single $\De$ being excluded,
(except if it is
an external square containing one of the source points $x\, $ and
$y$). One has the disjointness 
or hard core constraints $Y_{i}\cap Y_{j} = 0$ for $i \ne j$.\\ 
(ii) $b(Y) $ is computed from $a(Y) $ through
\eq
b(Y) \,=\, 
\sum{\!}'\,\, a^{l} (Y)\,\,,
\eqe
where the sum is over all assignments of large field regions
{\it included} in $Y$. 
This sum $\sum'$ is submitted to constraints (as indicated): We define
$\La_l(Y):=\La_l\cap Y\,=\,\bigcup_n \La_{l^n}(Y)\,,\,\,
\La_{l^n}(Y):=\La_{l^n} \cap Y$ and sum over the $s,l^n$-assignments
within $Y$ with the following restriction: 
For given $Y$ any assignment for which there exists some
$\De \in \La_{l^n}(Y)$ with 
\eq
{\rm dist} \ (\De , (\partial Y - \partial \La ) )\,\leq \,n\,M
\eqe
is forbidden, because otherwise our polymer would not contain the
whole of the large field block $\Ga ^e_a$ 
containing $\De$ and associated with $\La_l(Y)$. It is
also evident that it does contain this block if (138) does not hold
for any square from $\La_{l^n}(Y)$. 
With this definition of the amplitudes $b(Y)$ we now obtain
factorization: 
\eq
S_{2}(x,\,y) \,=\, { 
\sum\limits_{q, Y_{i} \atop
Y_{i }\cap Y_{j} =0 } b(Y_{1},\,x,\,y)
{1 \over (q-1)!}\prod_{i=2}^{q} b(Y_{i})
\over  
\sum\limits_{q, Y_{i}^{l} \atop
Y_{i }\cap Y_{j} =0 }
(1/q!)\prod_{i=1}^{q}  b(Y_{i}) }
\eqe

\subsection{ The Mayer Expansion and the Convergence Criterion}
(139) has now the form required
for the application of the Mayer expansion
in a standard way.
The hard core interaction between two clusters or polymers $X$, $Y$ 
is $V(X,Y)=0$ if $X \cap Y 
=0$, and $V(X,Y)=+\infty$ if $X \cap Y 
\not =0$, and the disjointness constraint for the polymers
can be replaced by the inclusion of
an interaction $e^{-V(Y_{i}, Y_{j})}$ between each pair of polymers.
A configuration $M$ is an ordered sequence of polymers.
We define $b^{T}(M) $ by
\eq
b^{T}(M)   = T(M)   ({1 \over q!} \prod_{i=1}^{q} b(Y_{i}))\,\,, 
\eqe
where the connectivity factor $T(M)$ is defined using connected Graphs
$G$ on $M$, by 
\eq
T(M):= \sum_{G \ {\rm connected \ on \ } M} \,\,\prod_{ij \in G} 
(e^{-V(X_{i}, X_{j})}-1)\,\,.
\eqe
Then we can divide by the vacuum functional to obtain 
\eq
S_{2} (x,\,y) \,=\, \sum_{M \ (x,\,y) -{\rm configuration}}
b^{T}(M)\,\,,
\eqe
where $M$ is a sequence of overlapping polymers $Y_{1}$, ..., $Y_{q}$,
the first of which contains the squares containing $x$ and $y$ and 
thus includes the factor 
$(\frac{1}{p_{reg}^2+m^2+ig\tau  \chi_{\La}})(x,y)\,$ from (73).
The sufficient condition for the convergence of (142) 
in the thermodynamic limit is well known: It is a particular bound on the sum
over all clusters, containing a fixed square
or point to break translation invariance [14,18,19]. We state it as\\
{\bf Proposition 10:}
\eq
\big| \sum_{Y, 0 \in Y } b(Y) e^{|Y|} \big| \le 1/2 
\eqe
for $N$ sufficiently large, uniformly in $\La$, $|Y|$ being the number
of squares in $Y$.\\[.1cm]
The fixed point is chosen to be 0 without restriction.
For $N$ large enough, (143) in fact holds if one replaces
the number $e$ in (143) by any other constant.
To deduce convergence of (142) under condition 
(143) requires to reorganize the connectivity factor $T(M)$
according to a tree formula. We can use again the basic forest 
formula (122) to obtain a symmetric sum over all trees.
We define 
\eq
v_{ij} = (e^{-V(X_{i}, X_{j})}-1) \ \ {\rm for} \ \ i \ne j \ \  .
\eqe 
We call $P$ the set of pairs $1 \le i < j \le n$. 
Expanding $ \prod_{(ij) \in P} (1 + v_{ij}) $ with (122) 
we get  another forest formula, on which we can read the
connectivity factor
\eq
 T(M) = \sum_{{\cal T}} \prod_{l\in {\cal T}} 
\bigl( v_{i_{l}j_{l}}\int_{0}^{1 } dh_{l}
\bigr)
\prod_{(ij)\not \in {\cal T}} (1 + h_{\cal T}(i,j) v_{ij}) \,\,,
\eqe
where $h_{{\cal T}}(i,j)$ is the $\inf$ of all parameters in the unique path
in the tree ${\cal T}$ joining $i$ to $j$.
This formula is then used e.g. like in [14,18,19] to derive the convergence
of (142). Remark that again every tree coefficient forces the
necessary overlaps and is bounded by 1.

It remains to prove Proposition 10. We do not give a first principles
proof here, but we do show how to sufficiently control {\it those}
contributions to the polymer amplitudes, which do {\it not}
 appear in analogous   
form in e.g. UV-regularized massive $\vp^4$-theory, since the latter is
clearly exposed in many reviews and textbooks, e.g. [14,19,22].  
Cluster expansion techniques are nowadays applied to much more
complicated situations than this, recently also with accent on a clear and 
systematic presentation [20,21].
The aspects not to be encountered in a $\vp^4$-treatment 
are analyzed in section 4.4. Here we reduce the proof
to certain bounds on functional derivatives generated by the links
of the second tree ${\cal T}_{2}-{\cal T}_{1}$ in (129). Because the amplitude
$b(Y)$ is given by a tree formula we will sum over all squares in $Y$ 
by following
the natural ordering of the tree, from the leaves towards the root, i.e.
the particular square containing 0. The factorial of the Cayley
theorem counting the number of (unordered) trees 
is compensated in the usual way by the symmetry factor $1/|Y|!$ that
one naturally gets
when summing over all positions of labeled squares [14,19]. Then the only
requirements to complete the proof of (143) are\\
(i) summable decay of the factor associated to each
tree link. This is obvious for the $\vep_{ij}$ links of 
${\cal   T}_{1}$, because these extend only over neighbours,
so have bounded range. For the tree links of ${\cal T}_{2}-{\cal T}_{1}$, it
follows from the decay of the corresponding kernels (125), see Lemmas 1,3. \\
(ii) A small factor for each tree link, or equivalently for each
square of $Y$. This will compensate in particular for the
combinatorial factors to choose which
term of the action to act on by the derivatives etc.
For tree links of ${\cal T}_{1}$ this small factor comes from the 
one associated to each of the large field squares, hence from Lemma 9.
Once a square is chosen large field we still have the choices
$l^1,\ldots l^{n_0}$, the value of $n_0$ depending on the distance of
the square from the boundary of $Y$. The sum over the $n$-values
converges (rapidly) due to (120).
For the tree links of ${\cal T}_{2}-{\cal T}_{1}$ the small factor
comes from the negative powers of $N$ generated
 at the ends of these links (``vertices''). These small
factors are described in more detail in the next section. 
Remark that  all types of small factors tend to zero as $N \to
\infty$.\\
 We note that the small factor per square 
should be there on taking into account the bound on the action 
as a net effect. (113) was derived before performing the cluster
expansion. Does it still hold once the interpolation parameters and
support restrictions are introduced? It does indeed, because support
restrictions do not cause any harm in the reasoning of Ch.3, because all
interpolated kernels are bounded in modulus by the modulus of their
noninterpolated versions (see (128)), and because the interpolated
versions of the operator $A$ still have real spectrum. Then one easily
realizes that all statements go through as before, in particular the
proof of Proposition 2 and of Lemma 7. A slightly more serious
modification of the action is caused by the use of the Cauchy formula 
below, it will be controlled by Lemmas 12 and 13.

\subsection{ The Outcome of the Derivatives}
With the tools previously developed we now want to show the existence
of the correlation functions in the thermodynamic limit. We have at
our disposal
exponentially decaying kernels, a suitable stability bound on the action
(Proposition 8), and we have arranged things such that derivatives
will produce a small factor  corresponding to the small coupling. 
As compared to a treatment of UV-regularized $\vp^4$. 
The main new features to be analysed are the following:\\
a) The action is nonlocal, and the covariance is interpolated twice.\\
b) There is a small/large field split, and thus small factors 
per derivative appear in various different forms.\\
c) The action is nonpolynomial, which implies in particular that terms
descended from the action by derivation may be rederived arbitrarily often.
       
The amplitudes of the polymers $Y$ are given as 
sums over trees (129) which are the factorized
contributions coming from the forest formula (122). 
When performing the $h$-derivatives those may
either apply to 
$d\mu _{\ga}(Y)$ or to
\eq 
(\frac{1}{p_{reg}^2+m^2+ig\tau \chi_Y})(x,y)\,\,
{\det}_3^{-N/2}(1+iA_s)
\,{\det}_2^{-N/2}(1+\frac{1}{1+iA_s}iA'')
\,\,e ^{1/2\,(\tau,\,\,\de C_{\ga}\,\tau)}\, \,.
\eqe
Here we went back to (73) (remembering that the term 
$e^{-R\int_{\bbbr^2-\La}\tau ^2}\,$ is now
absent, cf. Proposition 8).
In (146) the kernels from (125), which appear in $C_{\ga}$ and the action,
are to be replaced by their $h$-dependent 
versions. We write shortly $K(h)$ for 
$K(\{h_{{\cal T}_{2}-{\cal T}_{1}}\})$ and have (see (128),(129)...) 
\eq
K(h)(x,\,y)\,=\, \chi_Y(x)\,h^{{\cal T}_{1},{\cal T}_{2}}_{l}(h)\,K(x,y)
\chi_Y(y)\,\,.
\eqe
Application of derivatives with respect to 
$d\mu_{\ga}$ is evaluated by partial integration ([13], Chap. 9):
\eq
\partial_{h_{i} } \int d\mu_{\ga}^{Y} (h, \tau)\,\ldots \,=\,
\int d\mu_{\ga}^{Y} (h, \tau) \, \int_{x,y}   {\de \over \de \tau(x)}\, 
(\partial_{h_{i} } C_{\ga} (h))(x-y) 
 {\de \over \de \tau(y)  } \, \ldots
\eqe
In $C_{\ga}$  the kernels $S\,=\,{1 \over \sqrt{1+\pi} }$ are
interpolated. Thus $\partial_{h_{i} } C_{\ga} (h)$ is of the form
\eq
\partial_{h_{i} } C_{\ga} (h)\,=\,(\partial_{h_{i}} S(h))\,\hat{C_{\ga}}\,
S(h)\,+\, S(h)\,\hat{C_{\ga}}\,\partial_{h_{i}} S(h)\,\,.
\eqe
The supports of the derived kernels, i.e.
$\partial_{h_{i}} S(h)$, are by construction restricted to the 
two squares linked by the $h_{i}$ derivation [18], which  adds
a link to the previous tree.
Therefore the $\tau$ functional derivatives
are either directly localized in these squares - in the case where
$\partial_{h_{i}}$ applies to the first (second) kernel $S(h)$ in $C_{\ga}$,
and we consider the $ {\de \over \de \tau} $ derivative on the left (right),
or they are only essentially localized - when 
e.g. $\partial_{h_{i}}$ applies to the first (second) 
kernel $S(h)$ in $C_{\ga}$,
and we consider the $ {\de \over \de \tau} $ derivative on the right (left).
In the last case this means that 
the $ {\de \over \de \tau}$ functional derivative
is linked to its localization square via the  second (underived)
kernel $S(h)$, which  is  supported  over  
the polymer in question, see (128),(147).
It has exponential decay, so the links to squares distant from the
localization square rapidly decrease with distance.
Summing over them gives an additional factor $\sim 1/m^2\,$. Since
this tolerable deterioration of the bound per derivative is the only
effect of essential localization, we may forget about this difference 
from now on.

The  (${\cal T}_{2}-{\cal T}_{1}$)-$h$-derivatives can apply also to the
terms in (107). To roughly keep track of the combinatorial factors 
 involved we
note that any $h$-derivative may apply to any kernel in (146) ($\sim
10$ terms). If it applies to  the measure there appear two terms
with two functional
derivatives which again may apply to the action ($\sim 40$ terms).
Still one should note that the effect of these combinatorics is not 
very important since going through the terms in detail
(which we shall not do too explicitly) reveals that most of them give
much smaller (in $N$) contributions than the dominating ones. This is also 
true for the sum over the $l$-assignments: Large field contributions,
in particular for $n>1$, 
are tiny corrections due to (120). Therefore e.g. all
the contributions coming from the terms in $\de C_{\ga}$ are 
unimportant: They are only present when $\La_l \subset \ga$ is not empty. 
There is one more source of combinatoric increase of the number of
terms, namely due to the fact that the derivatives may also act on
terms produced by previous derivatives. 
For the polynomial part of the action this may only happen a few
times. But it needs special discussion when regarding the determinants.
%Since to any derivative 
%there is associated a link supported (essentially) in two squares
%this may happen very often only if one square is associated with many
%links. In this respect see the short discussion of local factorial
%principle below. So the net  effect is only in a certain deterioration
%of the bounds as regards the minimal $N$ required. To bound such
%effects optimally would require to regard the individual terms 
%more carefully than we do here, and in fact to perform a
%numerical analysis.   
So we will now go through the various contributions and comment how the
derivatives act on them. We can be short about\\[.1cm]
{\bf $\de C_{\ga}$}: In all terms we have the kernels $\sqrt{1+\pi}$,
which fall off as $\exp(-2m|x-y|)$. 
The contributions are listed in (70). When  
applying an $h$-derivative to $\de C_1(\ga)$ the small factor
in $N$ comes from $dist(\ga,\La-\Ga)\,\geq\,\ln N/m$. Due to the fall-off  
this gives a factor $\sim N^{-2}$. We may then e.g. write 
in the bound for the kernel 
\eq
\exp(-2m|x-y|)\,=\,\exp(-{5m \over 4}\,|x-y|)\,\exp(-{3m \over 4}\,|x-y|)
\eqe
and keep the first factor as a kernel with  exponential fall-off and bound
the second by $N^{-3/4}$ using the support restrictions. This  is then
the small factor per derivative. Note that we could also do without extracting 
this factor from (150), extracting it as a part of (120) instead. 
The same splitting (150) can be applied when the $h$-derivatives act
on $\de C_2(\ga)$. For $\de C_4(\ga)$ we may invoke support
restrictions to extract
$N^{-3/4}$ as above, additionally we get a factor of 
$\vep \sim N^{-2/5}$. The term  $\de C_3(\ga)$
does no more contribute due to the limit $R\to \infty$.
The same mechanism produces the small factors also, when we apply the
functional derivatives $\de/\de\tau$ instead of
$h$-derivatives. Remember the above remarks concerning essential
localization.  By the derivatives we also produce 
$\tau$-fields (essentially) localized 
in some square $\De$ (two fields per h-derivative, one per $\de/\de
\tau$-derivative). If the square $\De$ is in $\La_s$,
we have the choice to perform Gaussian integration or to bound the
contribution directly using (46) 
\eq
|\ldots\,\int_{\De} K_1(z-x)\,\tau(x) \,K_2(x-y)\,\ldots\,|\,
\leq\,\,O(1)N^{1/12}\,
|\ldots\,\,\sup_{x\in\De,} |K_1(z-x)\,K_2(x-y)|\,\,\ldots\,|\,\,.
\eqe
This is maybe the simplest way of doing. Note that in this case
we still can keep aside a factor of $N^{-3/4\,+\,2/12}\,<\,
N^{-1/2}$ per $h$-derivative. If the square is in $\La_{l^n}$,
the bound is achieved using (44),(45) and (120).
The above-mentioned rederivation of derived terms allows
to apply (at most) two $\de /\de \tau$ on an h-derived term
so that the factor  has to be distributed over three
derivatives leaving in this worst case  
$N^{-1/6}$ per derivative (without invoking large field suppresion).

Maybe we should also mention shortly the wellknown and well-solved 
local factorial problem. There is the
possibility that a large number of $\tau$-fields accumulate in a single
square $\De$, even when regarding only the polynomial part of the action, 
namely if the tree in question has a large coordination number 
$d$ at  that square: There are $d$ links of the type $l_{i,j_{\nu}}$,
$\nu =1,\ldots d$ in the tree, $i$ referring to $\De$. 
Then bounding the at most $2d$ $\,\tau$-fields in $\De \subset \La_l$ 
using (120) (and the Schwarz inequality) gives
\eq
[\int_{\De} \tau ^2 ] ^{d}\,e ^{-1/4\int_{\De}\tau ^2}\,\leq\,
4^d\, d!
\eqe
This is not tolerable as a bound when aiming to prove (143),
but the solution is in the fact that most of the $d$ squares associated  
to the links $l_{i,j_{\nu}}$ have to be at a large distance from $\De$ for
large $d$. Extracting a small fraction $\eta$ of the kernel decay 
we can isolate a  factor associated to $d >\!>1$, which is much
smaller than ${1 \over d!}$.
\footnote{It is of order $e ^{-\de \,d^{3/2}}$.} For a more 
thorough discussion of the point see [14,19] or also [1].\\[.2cm]
Now we regard the Fredholm determinants.
As compared to [1] we have to regard an inverted determinant. 
This is related to the fact that we regard a bosonic model,
and it  means that the sign cancellations appearing as a consequence 
of the Pauli principle which sometimes improve the convergence
properties are absent. 
The inverted determinants are raised to the power $N/2$. For shortness
we will {\it change the notation} for the rest of this
section and suppress this power assuming instead the operators 
$A_s\,,\ldots$ to
act in $\bigoplus_{k=1}^{N/2}{\cal L}^2(\La)$. We assume $N$ to be
even, otherwise we still would have to carry around a power $1/2$
(without consequence). This change entails  that we absorb a factor of
$N/2$ in $Tr$ as well.\\ 
We rewrite the product of the two Fredholm
determinants appearing in terms of a single one. This is possible,
since the interpolation 
acts equally on all $A$-operators. We have
\eq
{\det}_3^{-1}(1+iA_s)
\,{\det}_2^{-1}(1+\frac{1}{1+iA_s}iA'')\,=\,
{\det}^{-1}(1+iA) \,e ^{Tr\{ iA_s-1/2(iA_s)^2+{1 \over 1+iA_s}iA''\}}
\,.
\eqe
Since the $Tr$ of $A''$ multiplied by any power of $A_s$ vanishes, whereas
$Tr (A_s+A'')\,=\,TrA$, we may rewrite (153) as 
\eq
{\det}_2^{-1}(1+iA)\,\exp Tr\{-1/2(iA_s)^2\}
\,\,.
\eqe
The cluster derivatives acting on (154)
will be evaluated as Cauchy integrals
over suitable (large) contours. Similar reasoning has been used by
Iagolnitzer and Magnen [23] in a renormalization group analysis of
the Edwards model and earlier by Spencer in the analysis of the decay
of Bethe-Salpeter kernels [24]. To obtain useful bounds using this
method requires that the derivatives $\pa_{h_l} A\,$ are always 
small in norm. 
At this stage we therefore really need the whole
cascade of large field splittings from the previous chapter.
We have\\
{\bf Lemma 11:} Let $l\in {\cal T}_2 -{\cal T}_1$ be a link of the
cluster expansion joining two   squares $\De$, $\De'$ 
such  that  $\pa_{h_l} A\,=P_{\De '}\,A\,P_{\De}$. Then we have 
\eq
||\pa_{h_l}A||\,\leq\, O(1)N^{-5/12}\,\exp\{-m\, d_l\}\,\,,  
\eqe
if  $\De$ is a small field square.
Here we set $d_l\,=\,dist(\De,\De ')$.
If  $\De$ is a large field square in $\La_{l^n}$, we find
\eq
||\pa_{h_l}A||\,\leq\, O(1)\,N^{-1/2}(\int_{\De} \tau ^2)^{1/2}
\,\exp\{-m\, d_l\}\,\leq\, O(1)\,N^{{n \over 12}-{1 \over 2}-2n}\,\,.  
\eqe
\\[.1cm]
{\sl Proof:} 
The result is obtained in the same way as when proving Proposition
2, if $\De$ is a small field square.
If  $\De$ is in $\La_{l^n}$, the distance 
between  the squares is by our expansion rules larger than  ${2n \ln N
  \over m} $ which assures (156) through the decay of ${1 \over p^2+m^2}$
(remember in particular (44),(45),(62)). \qed\\[.1cm]
For shortness of notation we introduce
\eq
{\det}^{-1}(1+Q):={\det}^{-1}(1+iA)
\eqe
and first describe how the derivatives act on (157) instead of (154). 
Namely  we write
\eq
\pa_{h_1}\,\ldots\,\pa_{h_n}\,{\det}^{-1}(1+Q)\,=\,
\eqe
\[
\Bigl[\pa_{\al_1}\,\ldots\,\pa_{\al_n}\,
{\det}^{-1}(1+Q+\al_1\pa_{h_1}Q+\ldots+ \al_n\pa_{h_n}Q)\,\Bigr]_
{\al_1,\ldots,\al_n\,=\,0}
\]
We evaluate (158) by means of a Cauchy formula for the $n$
independent complex variables $\al_i$. The idea is to
regain the small factor per derivavtive and the distance decay by
choosing the $\al$-parameters sufficiently large. We note first that
${\det}^{-1}(1+Q+\al_1\pa_{h_1}Q+\ldots+ \al_n\pa_{h_n}Q)$ is analytic
in the $\al$-parameters, see Simon [11], as long as 
$1+Q+\al_1\pa_{h_1}Q+\ldots+ \al_n\pa_{h_n}Q$ has no 0 eigenvalues.
This restricts the maximal size of the $|\al_i|$. We choose 
the size of the $\al_l$-parameter corresponding to the link $l$ as follows:
\eq
R_l :=\,|\al_l|\,=\, N^{1 \over 6} \, e ^{{9 m \over 10}\, d_l}\,\,.
\eqe
\\
We now find\\
{\bf Lemma 12:} If the $\al_l$ are chosen according to (159) then 
\eq
||\sum_l \al_l\,\pa_{h_l}A||\,\leq\,O(1)\,N^{-1/4}
\eqe
{\sl Proof:} For the individual entries in the sum the bound 
follows on inspection. If the supports of the links 
(i.e. the pairs $\De$, $\De '\,$) are mutually
disjoint it stays true, since then the $\pa_{h_l}A\,$ are mutually
orthogonal. If they are not, we again employ the argument (see above (152)) 
that in this case the links corresponding to a large coordination
number $d$ in the tree have to grow longer and longer. 
In this case the sum may be performed using the remnant decay
$e ^{-{m \over 10 }d_l}$. \qed 
{\it Remark:} When proving the exponential decay of the two-point
function in the end of the paper we would like have exponential decay 
with mass $m$ up to corrections small with $N$ (without invoking the
analyticity improvement due to the UV cutoff). It may then be necessary
to use the {\it full} decay for {\it at most two} links 
\footnote{ if these links are indispensable to join via the tree the squares
  containing the points $x$ and $y$ in the external polymer $A(Y,x,y)$.} 
among those appearing at a branch point of the respective tree (see below,
proof of Theorem). Obviously this does not
change the norm bound (160) at all, since we may 
bound the sum in the same way keeping aside
a fraction of the decay for $d-2$ links only.\\[.1cm]
So we now evaluate (158) through
\eq
|\pa_{h_1}\,\ldots\,\pa_{h_n}\,{\det}^{-1}(1+Q)|\,=\,
|({1 \over 2\pi i})^n\,\int_{R_1\ldots R_n}\,
{1 \over \al_1^2 \ldots \al_n^2}\, {\det}^{-1}(1+Q+\sum_l \al_l
\pa_{h_l}Q)\,|
\eqe
\[
\,\leq\,
({1 \over 2\pi })^n\,{1 \over R_1,\ldots, R_n}\,\, \sup_{\al} 
|det^{-1}(1+Q+\sum_l \al_l \pa_{h_l}Q)\,|\,\,,
\]
where the $\sup$ is to be taken over the $\al_l$-parameters on the circles
$R_l$.
Thus we obtain indeed per derivative a factor 
\eq
{1 \over 2\pi}\,N^{-1/6}\,e ^{-{9 m \over 10}d_l}\,\,. 
\eqe
Before ending the discussion of how to evaluate derivatives 
acting on $\,\det\,$ we mention how we treat the 
$\de /\de \tau_m$-derivatives. In this case 
we choose (in modification of (159))
\eq
R_m(\tau)\,=\,N^{1/4}\,\,. 
\eqe
We thus  collect a smaller factor in $N$ from the $\tau$-derivative
because $\de /\de \tau$ annihilates a possibly large $\tau$-factor,
on the other hand we do not get a distance decay factor and need not do so,
because it is already present in the term ${\pa}_{h_l} S(h)$ which
accompanies $\de /\de \tau$ (see (149)).

Of course it remains to  give suitable bounds 
on the Fredholm determinants modified by the $\al$-parameters.
We have to remember that our true object of interest
is not ${\det}^{-1}(1+Q)$ but rather the subtracted determinant (154). 
First we note that we may still evaluate the derivatives acting on
(154) by introducing $\al$-parameters, on replacing as before
for a given choice of of $h$- and $\tau$-derivatives
\eq
A_{\al}:=\,A\,\to\,A\,+\,\sum_l \al_l \pa_{h_l}A \,+\,
\sum_m \al_m \de_{\tau_m} A
\,\, \mbox{ and similarly for } \,\, A_s\,,\,\,\, A'' \,\,. 
\eqe
So after bounding the Cauchy integrals  we have to bound 
\eq
\sup_{\al} |{\det}^{-1}_2(1+iA_{\al})\,
\exp \{{1 \over 2} Tr (\,A_{\al\,s})^2 \}\,|\,. 
\eqe
The task is to reproduce the bounds on the action from Ch.3 
on replacing $\,A \to A_{\al}\,$. Inspection shows that the proofs
of Propositions 2, Lemmas 6,7 and  part of Proposition 8 (as far as   
(117) is concerned) have to  be redone with  this modification on 
$A$. We collect our findings in\\
{\bf Lemma 13:} We assume that the kernels $A$ are restricted
to a given polymer $Y \subset \La$ of the cluster expansion. Then we have\\   
a) $\,||A_{\al \,s}||\,\leq\,O(1)N^{-1/4}\,$ (replacing (48))\\
b) $|\,{\det}^{-1}(1+{1 \over 1+iA_{\al\,s}}i A''_{\al})|\,\leq
\exp\{O(1)N^{-1/4}\int_{Y}(\tau ^2\,+\,1) \}\,$ (replacing (109))\\
c) $\,|Tr(A_{\al\,s}^3)|\,\leq \, N^{-1/4}\,\int_{Y\cap \La_s} \tau ^2$ 
(replacing (117))\\[.1cm]
{\sl Remark:} Note again that due to our change of notation a factor of $N/2$ 
has been absorbed in $Tr$ together with a corresponding change in
$\det$.\\[.1cm] 
{\sl Proof:} The proof of a) is trivial from Proposition 2 and Lemma 12.
As for b) we have to go again through the considerations leading from
(95) to (109). Since the reasoning is analogous, we will be rather
short. Introducing the quantities $B_{\al},\, D_{\al}$ as we did for 
$A_{\al}$ we find that (95) to (97) stay true also for complex $\al$.
The essential modification occurs in (101),(102): since the $\al_i$ are
complex we  find
\eq
i(\vp,\, A_{\al}''\vp)\,-\,i(\vp,\,A_{\al}''\vp)\,=\,-\, 
2\sum_{\al_l} Im\,\al_l \,(\vp,\, \pa_{h_l} A''\vp)\,-\,
2\sum_{\al_m} Im\,\al_m \,(\vp,\, \de_{\tau_m}A''\vp)
\eqe 
instead of $0$ for $\al\equiv 0$. 
Correspondingly we have to modify (102). The norm bound (160) 
then still implies
\eq
D_{\al}\,\leq \, O(1)\,N^{-1/4}\,\,.
\eqe
which is weaker than (105) but sufficient for us. In evaluating 
$Tr D_{\al-}^2$ we take into account the additional contribution too.
Since now
\eq
0\leq\,
D_{\al-}\,\leq \, (2+\eta)\,|\de\,A''|\,+\,
\sum_l 2\,R_l \,| \pa_{h_l} A''|\,+\,
\sum_m 2 \,R_m(\tau) \,|\de_{\tau_m}A''|
\,,\quad \eta <\!<1\eqe
it is straightforward to realize that we may bound
$TrD_{\al-}^2$  by
\eq
Tr D_{\al-}^2 \,\leq\, O(1)N^{-1/4}(\int_{\La\cap Y} \tau^2\,+\,1)\,\,.
\eqe
The first contribution is obtained similarly as in (108), 
it is quadratic in $\tau$, but we can keep aside a small factor. The
additional contribution is proportional to the number of squares touched
by $\de_{\tau}$-derivatives ($\leq\,|Y|\,$), thus it  is independent
of the size of
$\tau$. This ends the proof of b).\\
c) The proof is as in (116), (117).\qed
From Lemma 13
we now find that (113) (on restriction to $Y \subset \La$ and on using
interpolated kernels) is to be
replaced by 
\eq
Z^Y_{\ga} G^Y_{\ga,\al}(\tau)
\,\leq\, e ^{-{49 \over 100}\int_{\La_l\cap Y}\tau ^2}\,
e ^{ O(1)N^{-1/4}\int_{\La_s \cap Y}(\tau ^2\,+\,1)}\,\,.   
\eqe
So the large field suppresion stays unaltered and in the bound 
on the polymer amplitudes  there is at most a factor of $\sim
1\,+\,O(N^{-1/12})$
per  small field square from the action to beat (we could tolerate O(1)). 

Here we may end our discussion on the outcome of the derivatives. 
We have shown that we have a small factor 
$\sim N^{-1/6}$ per  derivative and factor of $e ^{-N^{1/8}}$  
per large field square. 
All links are through kernels decaying exponentially with mass $>m$.
This is sufficient to beat the factors $O(1)$ per square 
from the combinatoric choices 
\footnote{where we mentioned already that just taking the maximal
value gives a crude bound since most terms are much smaller than the
leading ones} and from the action. We pointed out that this is
sufficient for the proof of Proposition 10.

\subsection{ Exponential Decay of the Correlation Functions}
Now we have proven the existence of $S_2(x,y)$ in the TD limit we want
to proceed to the announced result on its exponential decay.\\
{\bf Theorem:} For $N>\! >1$ sufficiently large the inifinite volume 
two-point function decays exponentially 
\eq
|S_2(x,y)|\,\leq\, O(1)\,e^{-m'|x-y|}
\eqe
with 
\eq
m'\,=\,m(1\,+\,o(N^{1/10}))\,\,.
\eqe
{\sl Remarks:} O(1) is an $N$-independent positive number. The
estimate  on the
exponent of $N$ in (172) is of course not optimal. The proof goes through
without much change also for any $2n$-point function. Using the
effects of the UV-cutoff we could replace $m'$ by $m$.\\[.1cm]
{\sl Proof:}
The reasoning is very similar to that of [1] though somewhat simpler.
The point is now to realize that the convergence proof still works
when we put aside the decay factor appearing in (171). We may assume
$x$ and $y$ far apart. They both have to be contained in the same polymer
$A(Y,x,y)$, and we have to extract the decay factor when calculating its
amplitude. More specifically we shall extract it from
the sum over trees ${\cal T}_2$ in (129), 
where we first only  deal with those trees
${\cal T}$ for which ${\cal T}_1$ is empty, namely  
we first assume that $Y$ does not
contain large field squares, which is the dominant contribution.
Obviously the decay is associated with the factor 
\eq
\bigl[ \frac{1}{p^2+m^2+ig\tau} \bigr](x,y)\,=\,
\bigl[ \frac{1}{1\,+\,\frac{1}{p^2+m^2}\,ig\tau}\, \,\frac{1}{p^2+m^2} \bigr]
(x,y)\,
\eqe
which  appears in the external polymer.
The kernel $\frac{1}{p^2+m^2}$ is interpolated and thus in particular 
restricted in support to $Y$.
Let $\De_1$ and $\De_2$ be the squares in $Y$ containing $x$
and $y$. For given tree ${\cal T}$ there is a unique path in
${\cal T}$ connecting $\De_1$ and $\De_2$. We call it ${\cal T}'$,
noting that ${\cal T}'$
is a tree with coordination numbers $d_i= 2$, apart from
the ends, where they equal $1$. Its
complement in ${\cal T}$ will be called ${\cal T}''$. It has several
connected components in general. Each of these connected components
may be viewed as being rooted at some square 
(attached to links) from ${\cal T}'$. Keeping these squares fixed for
the moment and summing over the positions of the other squares in the various
connected components  of ${\cal T}''$ then provides us for these
connected components with the usual
polymer bound  (Proposition 10) sufficient for convergence. 
It remains to sum over the positions
of the squares in ${\cal T}'$ apart from $\De_1$ and $\De_2$, which
are sitting on the ends. For given positions of those squares we may
isolate a factor of
\eq
\vep^{|{\cal T}'|}\,\prod_{l'\in {\cal T}'}K_{l'}(x_{l'},y_{l'})\,\,.
\eqe
Here $\vep \,\sim\,o(N^{-1/10})$ is part of the small factor per
small field derivative, the other being used to beat the combinatoric
constants etc., see above. The kernels $K_{l'}(x_{l'},y_{l'})$ are those
generated by the derivatives of the expansion. They all fall off
exponentially with at least the rate of  $\frac{1}{p^2+m^2}$, so they
all may be bounded by the modulus of
$[\frac{1}{p^2+m^2}](x_{l'},y_{l'})$
up to a constant $\sim O(1)$, which we absorb in $\vep$. 
The coordinates $(x_{l'},y_{l'})$ are situated in the two squares
linked by $l'\in {\cal T}'$ and are to be integrated over those
squares.\footnote{apart from $x_{1'}=x$ and $y_{|{\cal T}'|}=y$ which
  are fixed} In Lemma 1 we showed that the kernel of 
$\frac{1}{p^2+m^2}$ is pointwise positive. From this we then obtain
easily that (174), when integrated over the intermediate squares 
and summed over their positions is bounded by
\eq
\vep^{|{\cal T}'|}\,
\bigl[ \bigl(\frac{1}{p^2+m^2}\bigr)^{|{\cal T}'|}
\bigr](x,y)
\eqe
(up to a constant $\sim O(1)$, which we absorb in $\vep$.)
Note that having split up the tree  ${\cal T}$ does not change the way
in which the sum over the trees is performed. We succeeded
in extracting  the factor (175) due to the fact that 
{\it two} squares in the external polymer are
fixed instead of only one as in Proposition 10. 
When summing over all possible values of $|Y|$ and using the polymer
bound (143) we now obtain a bound of the form
\eq
|S_2(x,y)|\,\leq\, O(1) \,\bigl[\frac{1}{p^2+m^2}\bigl(1\,+\,
\,\sum_{|{\cal T}'|}\vep^{|{\cal T}'|}\,
 (\frac{1}{p^2+m^2})^{|{\cal T}'|}\bigr)
\bigr](x,y)\,\,.
\eqe
Here the first term is the contribution for  $|Y|=2$ and where the 
single $h$- derivative applies to the second factor in (173). This is the
only case where it does not produce a factor $\,\leq \vep$.
Performing the geometric series in (176) now proves (171) on using
\eq
\bigl( \frac{1}{p^2+m^2-\vep\,}\bigr) (x,y)\leq\,
O(1)\,\exp \{ -(m-\vep/m)|x-y|\} \,\,.
\eqe
Finally we have to make sure that large field contributions do not
spoil our estimate. For this it suffices to note that in the large
field region we have at our disposal a factor of 
$\leq \exp(-N^{1/8})$ per square of $\Ga ^e$, half of which 
 may be put aside per each square of $\Ga_i^e$, on which  ends some 
$l'\in {\cal T}'$. Then we only have to note that this factor is much smaller
than the factor of  $\vep$ which we loose instead, and that the links
within $\Ga_i ^e$  are of short range. 
\qed\\[.3cm]
{\bf Acknowledgement:} The author is indebted to Jacques Magnen and 
Vincent Rivasseau for many helpful remarks. In particular the paper 
was initiated through several discussions with Jacques Magnen.
The important reference [25] was pointed out to me by K.Gawedzki.
\\[1.2cm]  
\noindent
{\bf References}
\begin{itemize}
\item[ [1] ]   Ch.Kopper, J.Magnen, V.Rivasseau:
Mass generation in the large $N$ Gross-Neveu-Model,
Commun.Math.Phys.{\bf 169}(1995)121-180.
\item[ [2] ] J.Zinn-Justin,  Quantum Field Theory and Critical
Phenomena, 3rd ed.,Clarendon Press,Oxford (1997).
\item[ [3] ] For a review see: 
F.A.Smirnov, Form Factors in Completely Integrable Models of Quantum
field Theory, World Scientific, Singapore(1992). Some important 
references are: A.Zamolodchikov, Al.B.Zamolodchikov:
Relativistic Factorized $S$-Matrix in two dimensions having $O(N)$
isotopic symmetry, Nucl.Phys.{\bf 133}(1978)525. 
M.Karowski, P.Weisz, Nucl.Phys.{\bf B139}(1978)455.
P.Hasenfratz, M.Maggiore, F.Niedermayer: 
The exact mass gap of the $O(3)$ and $O(4)$ $\,\si$-models in $d=2$, 
Phys.Lett.{\bf B245}(1990)522-528. P.Hasenfratz, F.Niedermayer:
The exact mass gap for the $O(N)$ $\,\si$-model for arbitrary $N\geq 3$
in $d=2$, Phys.Lett.{\bf  B245}(1990)529-534.
\item[ [4] ] K.Gawedzki, A.Kupiainen: Continuum Limit of the
  Hierarchical $O(N)$ Nonlinear $\si$-Model,
Commun.Math.Phys.{\bf 106}(1986)533-550 .  
\item[ [5] ] A.Pordt,  Th.Reiss:
On the renormalization group iteration of a two-dimensional
hierarchical nonlinear $\si$-model,
Annales de l'Institut Poincar{\'e} {\bf 55}(1991)545-587. 
\item[ [6] ] E.Br{\'e}zin, J.Le Guillou and J. Zinn-Justin:
Renormalization of the nonlinear $\si$ model in 
$2+\vep$ dimensions, Phys.Rev.{\bf D14}(1976)2615-2621.
E.Br{\'e}zin, J. Zinn-Justin: Spontaneous Breakdown of Continuous
Symmetries near two dimensions,
Phys.Rev.{\bf B14}(1976) 3110-31120.
\item[ [7] ] 
see e.g. S.Caracciolo, R.Edwards, A.Pelissetto and A.Sokal:
Asymptotic Scaling in the Two-dimensional $O(3)$ $\si$ Model at
Correlation Length $10^5$, Phys.Rev.Lett.{\bf 75}(1995),
1891-1894.
\item[ [8] ] A.Patrascioiu, E.Seiler: Super-Instantons 
and the Reliability of Perturbation Theory in Non-Abelian Models,
Phys.Rev.Lett.{\bf 74}(1995),1920-1923, and:
Nonuniformity of the $1/N$ Expansion for  $O(N)$ Models,
Nucl.Phys.{\bf B443}(1995)596.
\item[ [9] ]  
N.D. Mermin and H.Wagner: 
Absence of ferromagnetism and antiferromagnetism in one- and
two-dimensional isotropic Heisenberg models, Phys.Rev.Lett.{\bf
  17}(1966)1133-1136, 
N.D.Mermin: Absence of ordering in certain classical
systems, Journ.Math.Phys.
{\bf 8}(1967)1061-1064.
\item[ [10] ]
R.Dobrushin and S. Shlosman: Absence of breakdown of continuous
symmetries in two-dimensional models of statistical mechanics,
Commun.Math.Phys.{\bf 42}(1975)31-40. 
\item[ [11] ]   
E.Seiler: Schwinger functions for the Yukawa Model in two space 
time dimensions with space time cutoff, 
Commun.Math.Phys.{\bf 42} (1975) 163-182.\\
B.Simon: Trace Ideals and their Applications,
London Mathematical Society Lecture Note Series 35,
Cambridge Univ.Press (1979).\\
P.A.Faria da Veiga: Constructions de Mod{\`e}les non renormalisables en
Th{\'e}orie quantique des Champs, Thesis Ecole Polytechnique 1991.\\
J.Magnen and R.S{\'e}n{\'e}or: Yukawa Quantum Field Theory in three
Dimensions, Proc.3rd Int.Conf.on Collective Phenomena, Annals of the
New York Academy of Sciences {\bf 337}, New York 1980. 
\item[ [12] ] M.C.Reed in: Constructive Field Theory,
Proc. Erice 1973, Lecture Notes in Physics {\bf 25}, (1973)
\item[ [13] ] J.Glimm, A.Jaffe: Quantum Physics, Springer-Verlag, New
York 1987.
\item[ [14] ] D.Brydges in: Critical  Phenomena, Random Systems,
Gauge Theories, Proc. Les Houches 1984. North Holland (1986).
D.Brydges and Ph.A.Martin: Coulomb Systems at low densitiy,
preprint 1998.
\item[ [15] ] D.Brydges and T. Kennedy: Mayer Expansions and the
Hamilton-Jacobi Equation, Journ.Stat.Phys.{\bf 48} (1987) 19. 
\item[ [16] ] D. Brydges and H.T. Yau: 
Grad$\Phi$ Perturbations of massless Gaussian Fields,
Commun.Math.Phys.{\bf 129} (1990) 351. 
\item[ [17] ] D.Brydges and P.Federbush: A new Form of the Mayer
Expansion in Classical Statistical Mechanics, 
Journ.Math.Phys.{\bf 19} (1978) 2064. 
\item[ [18] ] A.Abdesselam and V.Rivasseau: Trees, Forests and
Jungles, a Botanical Garden for Cluster Expansions, 
Proceedings of the International Workshop on Constructive
Theory, Springer Verlag 1995, ed. V. Rivasseau.
\item[ [19] ] V.Rivasseau: From Perturbative to Constructive
Renormalization, Princeton University Press 1991.
\item[ [20] ] A.Abdesselam: Renormalisation Constructive Explicite, 
thesis Ecole Polytechnique 1997. 
\item[ [21] ] D.Brydges, J.Dimock and T.Hurd: 
Estimates on Renormalization Group 
Transformations, Univ.of Virginia preprint 1996, 
and: A non-Gaussian fixed point for $\phi^4$ in
$4-\vep$ dimensions, Univ.of Virginia preprint 1997, to appear in
Commun.Math.Phys.1998.  
\item[ [22] ] D.Brydges: Functional integrals and their Applications,
EPFL Lecture Notes, Lausanne 1992.
\item[ [23] ]  D.Iagolnitzer and J.Magnen:
Polymers in a Weak Random Potential in Dimension Four: Rigorous
Renormalization Group Analysis,
Commun.Math.Phys.{\bf 162} (1994),85-121.
\item[ [24] ] T.Spencer: The decay of the Bethe-Salpeter kernel in
$P(\vp)_2$ Quantum Field Theory, 
Commun.Math.Phys.{\bf 44} (1975),153-164.
\item[ [25] ] A.Kupiainen: 
On the $1/n$ Expansion, Commun.Math.Phys.{\bf 73} (1980),273-294.
\item[ [26] ] K.R.Ito, H.Tamura: 
$N$ Dependence of Upper Bounds of Crititcal Temperatures of $2D$
$O(N)$ Spin Models, Commun.Math.Phys.,
  this volume.
\end{itemize}
\end{document}